\documentclass[12pt, draftclsnofoot, onecolumn]{IEEEtran}
\ifCLASSINFOpdf

\else

\fi
\usepackage{mathrsfs}

\usepackage{amsmath}
\usepackage{ntheorem}

\usepackage{makeidx}  
\usepackage{algorithm}
\usepackage{algorithmic}
\usepackage{graphicx}
\usepackage{subfigure}
\usepackage{epstopdf}
\usepackage{bm}
\usepackage{cite}
\usepackage{stfloats}

\usepackage{amssymb}
\usepackage{graphicx}
\newtheorem{theorem}{Theorem}
\usepackage{url}

\usepackage{tabularx,booktabs}
\newcolumntype{C}{>{\centering\arraybackslash}X} 
\usepackage{lipsum}

\usepackage{makecell} 

\usepackage{graphicx}
\usepackage{color}

\newtheorem{lem}{Lemma}
\newtheorem{remark}{Remark}
\newtheorem{corollary}{Corollary}

\begin{document}
\title {Is RIS-Aided Massive MIMO Promising with ZF Detectors and Imperfect CSI?}
\author{Kangda Zhi, Cunhua Pan, Gui Zhou, Hong Ren, Maged Elkashlan, and Robert Schober, \IEEEmembership{Fellow, IEEE}\thanks{(Corresponding author: Cunhua Pan).
		 		
		K. Zhi, C. Pan, G. Zhou, M Elkashlan are with the School of Electronic Engineering and Computer Science at Queen Mary University of London, London E1 4NS, U.K. (e-mail: k.zhi, c.pan, g.zhou,  maged.elkashlan@qmul.ac.uk).
		
		H. Ren is with the National Mobile Communications Research Laboratory, Southeast University, Nanjing 210096, China. (hren@seu.edu.cn).

		R. Schober is with the Institute for Digital Communications, Friedrich-Alexander-University Erlangen-N\"{u}rnberg (FAU), Germany (e-mail: robert.schober@fau.de).	}

}

\maketitle

\begin{abstract}
	This paper provides a theoretical framework for understanding the performance of reconfigurable intelligent surface (RIS)-aided massive multiple-input multiple-output (MIMO) with zero-forcing (ZF) detectors under imperfect channel state information (CSI). We first propose a low-overhead minimum mean square error (MMSE) channel estimator, and then derive and analyze closed-form expressions for the uplink achievable rate. Our analytical results demonstrate that: $1)$ regardless of the RIS phase shift design, the rate of all users scales at least on the order of $\mathcal{O}\left(\log_2\left(MN\right)\right)$, where $M$ and $N$ are the numbers of antennas and reflecting elements, respectively; $2)$ by aligning the RIS phase shifts to one user, the rate of this user can at most scale on the order of $\mathcal{O}\left(\log_2\left(MN^2\right)\right)$; $3)$ either $M$ or the transmit power can be reduced inversely proportional to $N$, while maintaining a given rate. Furthermore, we propose two low-complexity majorization-minimization (MM)-based algorithms to optimize the sum user rate and the minimum user rate, respectively, where closed-form solutions are obtained in each iteration. Finally, simulation results validate all derived analytical results. Our simulation results also show that the maximum sum rate can be closely approached by simply aligning the RIS phase shifts to an arbitrary user.
\end{abstract}


\IEEEpeerreviewmaketitle
\section{Introduction}

Massive multiple-input multiple-output (MIMO) has been widely recognized as a cornerstone technology for fifth-generation (5G) and beyond wireless communications\cite{lu2014overview,bjornson2017massive,ngo2013energy,zhang2014power,xiao2020uav,ozdogan2019massive,liu2021cellfree}. Thanks to their spatial multiplexing gains, massive MIMO systems can simultaneously provide high quality of service for multiple users on the same time-frequency resource. Massive MIMO also has some other appealing properties, e.g., the transmit power can be reduced inversely proportional to the number of antennas without sacrificing the achievable rate.

However, conventional massive MIMO still has some drawbacks. The first one is the blockage problem. Due to the complex environment and user mobility, communication links may be blocked, in which case the channel strength could be severely degraded. Another problem is the high cost and energy consumption of the active radio-frequency (RF) chains. Massive MIMO commonly employs hundreds of antennas, each of which will be connected to a RF chain. Hence, this system incurs high hardware cost and energy consumption.

The recently developed technology of reconfigurable intelligent surfaces (RISs)\cite{di2020smart,pan2020intelligent,wu2019intelligent,pan2020multicell,huang2019reconfigu,yuxianghao2020robost}, also referred to as intelligent reflecting surfaces (IRSs), is a promising solution for tackling the above two issues in massive MIMO systems. On the one hand, since the RIS is a small, thin and light surface, it can be flexibly deployed at a carefully selected location with a favorable propagation environment. Therefore, RISs enable additional high-quality communication paths to overcome the blockage problem.
On the other hand, RISs are comprised of low-cost passive reflecting elements, which are much cheaper than active RF chains. Therefore, it is envisioned that RISs are beneficial for improving the energy efficiency of conventional massive MIMO systems.

Due to these appealing features, RIS-aided massive MIMO has gained growing research interests with many activities, focusing on various applications and different perspectives, such as
channel estimation \cite{he2019cascaded}, dual-polarized transmission\cite{de2021polarize},  millimeter wave (mmWave) communications\cite{wang2021massiveMIMO}, hardware impairments\cite{papazafeiropoulos2021intelligent}, multi-RISs co-design\cite{mei2021multibeam}, cell-free systems\cite{van2021reconfigurable}, antenna selection\cite{he2020reconfigurable}, and power scaling law analysis \cite{bijoson2020nearField,zhi2020power,zhi2021twotimescale}.

To fully understand the potential of RISs, it is essential to draw theoretical insights from information-theoretical expressions, which rigorously demonstrate the impact of the various system parameters. Fundamental information-theoretical expressions for conventional massive MIMO systems have been provided in, e.g., \cite{bjornson2017massive,ngo2013energy,zhang2014power}. It was shown that the achievable rate of conventional massive MIMO systems with $M$ antennas scales on the order of $\mathcal{O}\left(\log_2\left(M\right)\right)$. 
This naturally raises the question what the corresponding scaling law for massive MIMO systems after the integration of RISs is.
To answer this question, explicitly analytical rate expressions are required. 
It has already been shown that in RIS-aided single-user systems with $N$ reflecting elements, the achievable rate could scale as $\mathcal{O}\left(\log_2\left(N^2\right)\right)$\cite{wu2019intelligent,han2019large}, or even $\mathcal{O}\left(\log_2\left(N^4\right)\right)$\cite{9060923} if two RISs cooperate. Similar scaling orders were also reported for some other RIS-aided communication scenarios, such as the RIS-aided relay\cite{kang2021irs}, RIS with scattering parameter analysis \cite{shen2021parameter}, and RISs with hardware impairments\cite{qian2021scaling,xing2021hw}.  However, these works focused on the simple single-user case, and cannot be easily generalized to multi-user systems.

In fact, it is challenging to provide an insightful analysis for the rate scaling order of RIS-aided multi-user systems. This is because the resulting signal-to-interference-plus-noise ratio (SINR) expressions are more complicated and more involved than the interference-free signal-to-noise ratio (SNR) expressions for single-user systems, and also because the optimal RISs passive beamforming vectors cannot be given in closed form in case of multiple users. Some initial results were provided in \cite{zhi2020power} and \cite{zhi2021twotimescale} by considering RIS-aided massive MIMO with simple maximal ratio combining (MRC). For uncorrelated Rayleigh fading channels, it was proved that the achievable rate scales only as $\mathcal{O}\left(\log_2\left(1\right)\right)$ with respect to $N$. This is due to the severe multi-user interference, since the common RIS-base station (BS) channel is used by all users. To tackle this issue, most recently, the authors in \cite{zhi2021ergodic} firstly revealed that a rate scaling order $\mathcal{O}\left(\log_2\left(MN\right)\right)$ is achievable with zero-forcing (ZF), which demonstrates the huge potential of ZF detectors in RIS-aided massive MIMO systems.

However, there are two main limitations in \cite{zhi2021ergodic}. Firstly, ideal channel state information (CSI) of the aggregated channel including the superimposition of the direct channel and the reflected channel, was assumed. Secondly, the authors in \cite{zhi2021ergodic} only considered some initial performance analysis and RIS phase shift optimization, which lacks further insightful analysis. By contrast, this work aims to provide an analytical framework to gain an in-depth analysis for the performance of RIS-aided massive MIMO systems with ZF detectors under the realistic assumption of imperfect CSI.

Specifically, in this work, we first propose a low-overhead channel estimation scheme, in which the required pilot length is independent of $N$. We next perform a comprehensive theoretical analysis to reveal the explicit rate scaling order and answer the fundamental question whether the RIS-aided massive MIMO with ZF detectors is promising or not. Finally, based on majorization-minimization (MM) algorithms, we respectively optimize the RIS phase shifts to maximize the sum user rate and the minimum user rate. The detailed contributions are summarized as follows.

	 \itshape {1) Low-overhead channel estimation:}  \upshape We first propose a minimum mean square error (MMSE)-based method to estimate the aggregated channel in the systems, which is a superimposition of cascaded RIS channels and the direct channels. The length of pilots only needs to be no smaller than the number of users. We also analyze the impacts of various system parameters on the mean square error (MSE).
	
	 \itshape {2) Reveal rate scaling orders:}  \upshape We derive the closed-form ergodic rate expression and its insightful lower and upper bounds. The lower bound  shows that the data rates of all users are guaranteed to be on the order of $\mathcal{O}\left(\log_2\left(MN\right)\right)$, regardless of the RIS phase shift design. The upper bound  shows that the data rate of a specific user can be on the order of  $\mathcal{O}\left(\log_2\left(MN^2\right)\right)$, if the RIS phase shift is designed to align its beamforming to that user. We also demonstrate that these two analytical results are robust to RIS phase shift quantization errors.
	
	 \itshape {3) Answer the question whether the considered system is promising or not:} \upshape Based on the analytical results, we prove that RIS-aided massive MIMO systems with ZF detector are promising for three applications. It can provide ultra-high network throughput according to the high data rate scaling order for all users; it can help reduce $M$ inversely proportional to $N$ without sacrificing the data rate, which helps avoid the power hungry RF chains and is promising for green communications; it can help all users communicate with small transmit power, inversely proportional to $N$, which is promising for IoT applications.
	
	 \itshape {4) Low-complexity RIS optimization:} \upshape We design the RIS phase shifts to maximize the sum user rate and minimum user rate, based on the MM algorithm with closed-form solution in each iteration. We also show that aligning RIS phase shifts to an arbitrary user is an effective heuristic approach for maximizing the sum user rate. In addition, we demonstrate that maximizing the sum rate can also ensure a high minimum user date.

The rest of this paper is organized as follows. Section \ref{section1} describes the system and channel model. Section \ref{section2} proposes the MMSE channel estimation scheme. Section \ref{section3} theoretically proves that RIS-aided massive MIMO is promising with ZF detectors. Section \ref{section4} proposes the MM algorithm for solving the sum rate and minimum user rate maximization problems. Section \ref{section5} provides extensive simulations to verify the the correctness of analytical results and the effectiveness of proposed optimization algorithms. Finally, Section \ref{section6} concludes this work.

\emph{Notations}:  Boldface lower case and upper case letters denote the vectors and matrices, respectively. The inverse, conjugate transpose, conjugate and transpose of matrix $\bf X$ are denoted by ${\bf X}^{-1}$, ${\bf X}^H$, ${\bf X}^*$, ${\bf X}^T$, respectively. The $(m,n)$-th and $(m,m)$-th elements of the matrix are represented by $\left[{\bf X}\right]_{(m,n)}$ and $\left[{\bf X}\right]_{mm}$.
$\mathbf{X}\succ\mathbf{0}$ and $\mathbf{X}\succeq\mathbf{0}$ respectively denote that $\mathbf{X}$ is definite positive and semi-positive. $\mathcal{O}$ denotes the standard big-O notation. $\lambda_{\max }(\mathbf{X})$ and $\angle\mathbf{X}$ denote the maximal eigenvalue and the phase of matrix $\mathbf{X}$. $\mathbb{E}\{\cdot\}$ and $\mathrm{Cov}\{\cdot\}$ denote the mean and covariance operators.

\section{System and Channel Model}\label{section1}
\begin{figure}
	\setlength{\abovecaptionskip}{0pt}
	\setlength{\belowcaptionskip}{-20pt}
	\centering
	\includegraphics[width= 0.4\textwidth]{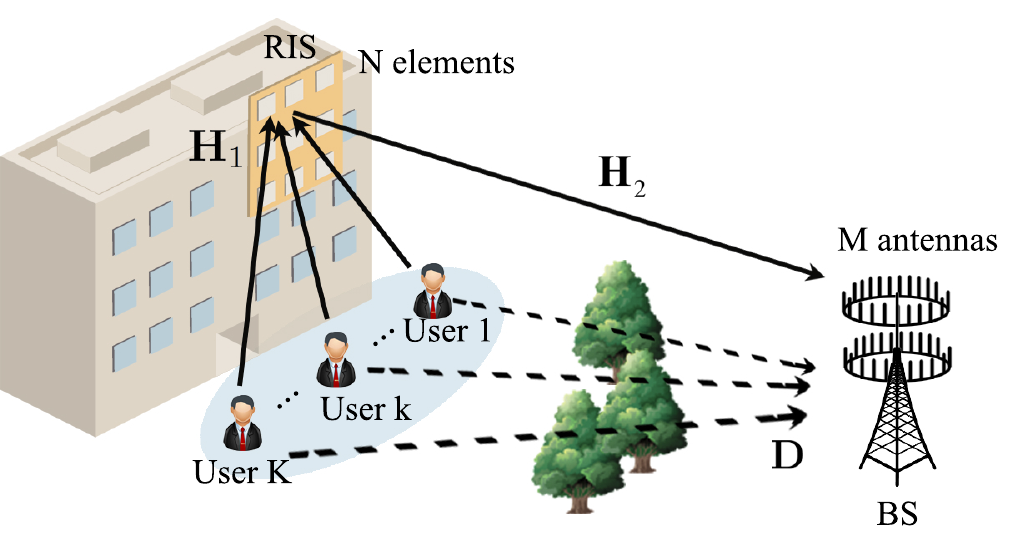}
	\DeclareGraphicsExtensions.
	\caption{Massive MIMO systems assisted by an RIS.}
	\label{figure1}
	\vspace{-10pt}
\end{figure}

As shown in Fig. \ref{figure1}, the uplink transmission of an RIS-assisted massive MIMO system is considered. 
The considered system consists of $K$ users with a single antenna, a BS with $M>K$ antennas, and an RIS with $N$ reflecting elements. Besides, we assume a quasi-static channel model with each channel coherence interval (CCI) spanning $ \tau_c $ time slots. In each CCI, we denote the instantaneous channel between the users and the RIS, and that between the RIS and the BS as $\mathbf{H}_1 \in \mathbb{C}^{N\times K}$ and $\mathbf{H}_2  \in \mathbb{C}^{M\times N} $, respectively. Then, the cascaded user-RIS-BS channel is  $\mathbf{G}=\mathbf{H}_2 \mathbf{\Phi} \mathbf{H}_1$, where $\mathbf{\Phi}=\mathrm{diag}\left\{e^{j\theta_1},\ldots,e^{j\theta_N}\right\}$ is the RIS phase shift matrix. Meanwhile, the direct channels between the users and the BS are denoted as $\mathbf{D} \in \mathbb{C}^{M\times K}$. Finally, in each CCI, the instantaneous 
\itshape {aggregated}  \upshape 
 channels from the users to the BS are given by $\mathbf{Q} = \mathbf{G} + \mathbf{D} \in \mathbb{C}^{M\times K}$.

It has been shown in \cite{zhi2021twotimescale} that it is better to place an RIS close to the users rather than close to the BS in the massive MIMO systems. Therefore, in this paper, we assume that the RIS is deployed on the facade of a tall building in the proximity of the users, as illustrated in Fig. \ref{figure1}. Since the RIS has a certain height and is close to the users, the user-RIS channels $\mathbf{H}_1$ would be line-of-sight (LoS) dominant. For analytical tractability, we assume that the user-RIS channels are purely LoS as follows
\begin{align}\label{H1}
\mathbf{H}_{1}=\left[\sqrt{\alpha_{1}} \;\overline{\mathbf{h}}_{1}, \ldots, \sqrt{\alpha_{K}} \;\overline{\mathbf{h}}_{K}\right],
\end{align}
where $\alpha_k, \forall k$ is the large-scale path loss factor for user $k$, and $\overline{\mathbf{h}}_k \in \mathbb{C}^{N\times 1}$ is the deterministic LoS channel between user $k$ and the RIS.

Since the RIS is installed close to the users, it may be located far away from the BS. Therefore, both LoS and non-LoS (NLoS) transmission paths would exist in $\mathbf{H}_2$.
As a result, we characterize the RIS-BS channel by Rician fading, which is expressed as
\begin{align}\label{H2}
\mathbf{H}_{2}=\sqrt{{\beta}/{(\delta+1)}}(\sqrt{\delta} \;\overline{\mathbf{H}}_{2}+\widetilde{\mathbf{H}}_{2}),
\end{align}
where $\beta$ is the path loss factor, and $\delta$ is the Rician factor which represents the ratio between the power of LoS component $\overline{\mathbf{H}}_2$ and the power of NLoS component $\widetilde{\mathbf{H}}_2$. The elements of $\widetilde{\mathbf{H}}_2$ are independent and identically distributed (i.i.d.) complex Gaussian random variables with zero mean and unit variance.  For a rich-scattering environment, we can assume $\delta\to 0$ and then the RIS-BS channel reduces to a Rayleigh fading channel containing only NLoS paths. For a scattering-free environment, we have $\delta\to\infty$ and then the RIS-BS channel is purely LoS.

Finally, since the users might be located far away from the BS, and rich scatterers (trees, cars, buildings and so on) are distributed on the ground, we assume that the channels between the users and the BS are Rayleigh fading \cite{han2019large}. Thus, we have
\begin{align}
\mathbf{D}\triangleq[\mathbf{d}_1,\ldots,\mathbf{d}_K]=\widetilde{\mathbf{D}} \mathbf{\Omega}^{1 / 2},
\end{align}
where $ \mathbf{d}_k =\sqrt{\gamma_k}\widetilde{\mathbf{d}}_k$ is the channel between user $ k $ and the BS with large-scale fading coefficient $ \gamma_k $ and small-scale fading vector $ \widetilde{\mathbf{d}}_k $ comprised of i.i.d.  complex Gaussian random variables with zero mean and unit variance. Here, $ \mathbf{\Omega}= \mathrm{diag} \left\{\gamma_1,\ldots,\gamma_K\right\} $ and  $ \widetilde{\mathbf{D}}=[\widetilde{\bf d}_1,\ldots,\widetilde{\bf d}_K]. $

We adopt the two-dimensional uniform rectangular array (URA) to model the LoS channels\cite{wu2019intelligent}. For an $L\times 1$ LoS channel $\mathbf{a}_{L}$, we first decompose $L$ into two closest integers $L_x$ and $L_y$, where $1\leq L_x\leq L_y, L_x\times L_y = L$. Then, 
the $l$-th element of $\mathbf{a}_{L}$ is given by
\begin{align}\label{USPA}
\begin{array}{l}
\left[\mathbf{a}_{L}\left(\vartheta^{a}, \vartheta^{e}\right)\right]_{{l}}=\exp \left\{j 2 \pi \frac{d}{\lambda}
\left(  \lfloor{(l-1)} /{L_y}\rfloor  \sin \vartheta^{e} \sin \vartheta^{a}+({(l-1)} \bmod {L_y})  \cos \vartheta^{e}\right)\right\},
\end{array}
\end{align}
where $\vartheta^{a}$ and $\vartheta^{e}$ denote the azimuth and elevation angles of arrival (AoA) or corresponding angles of departure (AoD). Based on (\ref{USPA}), it can be shown that $ \mathbf{a}_{L}^H\left(\vartheta^{a}, \vartheta^{e}\right) \mathbf{a}_{L}\left(\vartheta^{a}, \vartheta^{e}\right) =L$. Then, we can express that
\begin{align}\label{hk}
\overline{\mathbf{h}}_{k} &=\mathbf{a}_{N}\left(\varphi_{k r}^{a}, \varphi_{k r}^{e}\right), 1\leq k\leq K,\\\label{hk2}
\overline{\mathbf{H}}_{2} &\triangleq  \mathbf{a}_{M}  \mathbf{a}_{N}^{H}=\mathbf{a}_{M}\left(\phi_{r}^{a}, \phi_{r}^{e}\right) \mathbf{a}_{N}^{H}\left(\varphi_{t}^{a}, \varphi_{t}^{e}\right) .
\end{align}

\section{Channel Estimation}\label{section2}
To design the ZF detector, the channels are estimated by the BS using a pilot-based method. For conventional massive MIMO systems, only the $M\times K$ direct channel $\mathbf{D}$ needs to be estimated, and the minimum pilot sequence length is $\tau=K$. In RIS-aided massive MIMO systems, 
the required pilot overhead can be prohibitive due to the extremely large channel dimension of $ M\times N $ in the RIS-BS link. To reduce the pilot overhead, we only estimate the aggregated channel $\mathbf{Q}\in \mathbb{C}^{M\times K}$, for which the minimum pilot sequence length is still $\tau=K$, which is the same  as for conventional massive MIMO systems.

Specifically, in each CCI, the $K$ users are assigned mutually orthogonal pilot sequences with length $\tau\geq K$. The pilot sequence of user $k$ is denoted by $\mathbf{s}_k\in \mathbb{C}^{\tau\times 1}$. Let $\mathbf{S}=[\mathbf{s}_1,\ldots,\mathbf{s}_K]$, where $\mathbf{S}^H\mathbf{S}=\mathbf{I}_K$ due to the orthogonality. Then, at the beginning of each CCI, $\tau$ time slots are used for the $K$ users to transmit the pilot signal $\mathbf{S}$ to the BS. The received $M\times \tau$ pilot signal at the BS can be given by
$ \mathbf{Y}_p = \sqrt{\tau p} \mathbf{Q}\mathbf{S}^H + \mathbf{N}$,
where $\tau p$ is the transmitted pilot power of each user, and $\mathbf{N}$ is the noise matrix whose elements are i.i.d. Gaussian variables following $\mathcal{CN}(0,\sigma^{2})$. Then, we can obtain the observation vector for the channel of user $k$ by multiplying the term $\frac{1}{\sqrt{\tau p}} \mathbf{s}_k$ to $\mathbf{Y}_p$, as follows
\begin{align}\label{observation_vector}
\begin{array}{l}
	\mathbf{y}_p^k = \frac{1}{\sqrt{\tau p}} \mathbf{Y}_p \mathbf{s}_k = \mathbf{q}_k + \frac{1}{\sqrt{\tau p}} \mathbf{N}\mathbf{s}_k,
\end{array}
\end{align}
where $\mathbf{q}_k$, the $k$-th column of $\mathbf{Q}$, denotes the aggregated channel of user $k$.

\begin{lem}\label{lemma1}
Channel $\mathbf{q}_k$ and noise $\frac{1}{\sqrt{\tau p}} \mathbf{N}\mathbf{s}_k$ in (\ref{observation_vector}) are complex Gaussian distributed, where
$\mathbf{q}_k \sim \mathcal{CN}( \sqrt{\frac{\alpha_k\beta\delta}{\delta+1}}\overline{\mathbf{H}}_2\mathbf{\Phi}\overline{\mathbf{h}}_k,  \left( N \frac{\alpha_k\beta}{\delta+1} + \gamma_k\right)\mathbf{I}_M   )$, and $\frac{1}{\sqrt{\tau p}} \mathbf{N}\mathbf{s}_k \sim \mathcal{CN}   ({\mathbf{0}},   \frac{\sigma^2}{\tau p} \mathbf{I}_M  )$.
\end{lem}

\itshape {Proof:}  \upshape Please refer to Appendix \ref{App_1}. \hfill $\blacksquare$

From Lemma \ref{lemma1}, it is seen that the considered channel is still Gaussian distributed as conventional massive MIMO systems\cite[Eq. (1)]{ozdogan2019massive}, but with the different mean and variance. Therefore, we can still apply the well-used MMSE estimator to obtain the channel estimate of $\mathbf{q}_k$.

\begin{theorem}
The MMSE estimate of channel $\mathbf{q}_k$ is given by
\begin{align}\label{estimate}
\begin{array}{l}
\hat{\mathbf{q}}_{k}=\sqrt{\frac{\alpha_{k} \beta \delta}{\delta+1}} \overline{\mathbf{H}}_{2} \mathbf{\Phi} \overline{\mathbf{h}}_{k}+   \kappa_k  \left(\sqrt{\frac{\alpha_{k} \beta}{\delta+1}} \widetilde{\mathbf{H}}_{2} \mathbf{\Phi} \overline{\mathbf{h}}_{k}+\mathbf{d}_{k}+\frac{1}{\sqrt{\tau p}} \mathbf{N s}_{k}\right),
\end{array}
\end{align}
where $\kappa_k=\frac{N \frac{\alpha_{k} \beta}{\delta+1}+\gamma_{k}}{N \frac{\alpha_{k} \beta}{\delta+1}+\gamma_{k}+\frac{\sigma^{2}}{\tau p}} \in \left(0,1\right)$.
Denote the estimation error as $\mathbf{e}_{k}=\mathbf{q}_{k}-\hat{\mathbf{q}}_{k}$, where the error $\mathbf{e}_k$ is independent of the estimate $\hat{\mathbf{q}}_k$. Then, the MSE matrix for the channel estimation is
\begin{align}\label{MSE_matrix}
\mathbf{M S E}_{k}&=\mathbb{E}\left\{\mathbf{e}_{k} \mathbf{e}_{k}^{H}\right\}
=\frac{1}{\frac{1}{N \frac{\alpha_{k} \beta}{\delta+1}+\gamma_{k}}+\frac{\tau p}{\sigma^{2}}} \mathbf{I}_{M}
\triangleq \epsilon_k \mathbf{I}_{M}.
\end{align}

\end{theorem}

\itshape {Proof:}  \upshape Please refer to Appendix \ref{App_2}.  \hfill $\blacksquare$

Based on (\ref{MSE_matrix}), the MSE can be calculated as $\mathrm{MSE}_k = \mathrm{Tr}\left\{\mathbf{MSE}_k\right\} = \frac{M}{\frac{1}{N \frac{\alpha_{k} \beta}{\delta+1}+\gamma_{k}}+\frac{\tau p}{\sigma^{2}}}$. Clearly, the MSE is a decreasing function of $\tau$, $p$, and $\delta$, but an increasing function of $M$, $N$, $\alpha_k$, $\beta$, $\gamma_{k}$, and $\sigma^2$. This is because $\frac{\tau p}{\sigma^2}$ represents the pilot SNR, and increasing its value improves the estimation quality. $\delta$ is the Rician factor, and increasing its value makes the RIS-aided channels more deterministic and therefore decreases the estimation error. Also, the increase of $N$ introduces more communication paths between the users and the BS, which also increases the estimation error. 

Note that in the absent of the RIS (i.e., $\alpha_{k}=\beta=0, \forall k$) or for a purely LoS RIS-BS channel ($\delta\to\infty$), the MSE matrix in (\ref{MSE_matrix}) reduces to $\mathbf{M S E}_{k}=\frac{\gamma_{k}}{{1}+\frac{\tau p}{\sigma^{2}} \gamma_{k} } \mathbf{I}_{M}$, which is the same as for conventional massive MIMO systems \cite{ngo2013energy}. 
Let $\hat{\mathbf{Q}}=\left[\hat{\mathbf{q}}_{1}, \ldots, \hat{\mathbf{q}}_{K}\right]$ denote the estimated aggregated channel of the $K$ users. Then, based on (\ref{estimate}), we have
\begin{align}\label{hat_Q}
\begin{array}{l}
\hat{\mathbf{Q}}=\sqrt{\frac{\beta \delta}{\delta+1}} \overline{\mathbf{H}}_{2} \mathbf{\Phi} \mathbf{H}_{1}+\sqrt{\frac{\beta}{\delta+1}} \widetilde{\mathbf{H}}_{2} \mathbf{\Phi} \mathbf{H}_{1} \mathbf{\Upsilon}+\widetilde{\mathbf{D}} \mathbf{\Omega}^{1 / 2} \mathbf{\Upsilon}+\frac{1}{\sqrt{\tau p}} \mathbf{N S} \boldsymbol{\Upsilon},
\end{array}
\end{align}
where $\mathbf{\Upsilon}=\operatorname{diag}\left\{\kappa_{1}, \ldots, \kappa_{K}\right\}$. 

\section{Ergodic Rate Analysis}\label{section3}
In the transmission phase, the $K$ users transmit symbols $\mathbf{x}=[x_1,...,x_K]^{T}$ where $ \mathbf{x}\sim\mathcal{CN}(\mathbf{0},\mathbf{I}_{K}) $, and the received signal at the BS can be expressed as
\begin{align}
\mathbf{y} = \sqrt{p}\mathbf{Q}\mathbf{x} + \mathbf{n}
= \sqrt{p}\hat{\mathbf{Q}}\mathbf{x} + \sqrt{p}\mathbf{\mathcal{E}}\mathbf{x} + \mathbf{n}, 
\end{align}
where $\mathbf{n}\sim \mathcal{CN}\left(\mathbf{0},\sigma^2\mathbf{I}_M\right)$ and $\mathbf{\mathcal{E}}\triangleq\left[\mathbf{e}_{1}, \ldots, \mathbf{e}_{K}\right]=\mathbf{Q} - \hat{\mathbf{Q}}$. To eliminate the multi-user interference, the BS adopts the linear ZF detectors $\mathbf{A} = \hat{\mathbf{Q}}(\hat{\mathbf{Q}}^H\hat{\mathbf{Q}})^{-1}=\left[\mathbf{a}_1,\ldots,\mathbf{a}_K\right]$, which leads to $\mathbf{A}^H \hat{\mathbf{Q}} = \mathbf{I}_K$. Then, in each CCI, the BS detects the received signal as follows
\begin{align}
\mathbf{r} = \mathbf{A}^H \mathbf{y} = \sqrt{p}\mathbf{x} + \sqrt{p}\mathbf{A}^H\mathbf{\mathcal{E}}\mathbf{x} + \mathbf{A}^H\mathbf{n},
\end{align}
whose $k$-th entry can be further expressed as
\begin{align}\label{r_k}
	r_k = \sqrt{p}x_k + \sqrt{p} \sum\nolimits_{i=1}^{K} \mathbf{a}_k^H \mathbf{e}_i  x_i + \mathbf{a}_k^H \mathbf{n}.
\end{align}

\subsection{Derivatives of the Achievable Rate}
Based on (\ref{r_k}), the accurate ergodic rate of user $k$ can now be given by
\begin{align}\label{exact_rate}
\begin{array}{l}
R_k =  \tau^{o} \;\mathbb{E} \left\{\log_2\left(1 + \frac{p}{p  \sum_{i=1}^{K}  \left|\mathbf{a}_k^H \mathbf{e}_i \right|^2
	+    \sigma^2   \left\|  \mathbf{a}_k^H \right\|^2   }\right)\right\},
\end{array}
\end{align}
where a factor $ \tau^{o} \triangleq\frac{\tau_{c}-\tau}{\tau_c}$ captures the rate loss caused by pilot overhead,
and the expectation is taken over random channel components in $\hat{\mathbf{Q}}$.
It is difficult to derive an exact expression of (\ref{exact_rate}) due to the expectation operator before the logarithm symbol. Since the function $f\left( x\right)= \log_2\left(1+1/x\right)$ is convex of $x$, we utilize the Jensen's inequality to obtain the following lower bound
\begin{align}\label{positive_rate}
R_{k} \geq \underline{R_k} \left(\mathbf{\Phi}\right)&\overset{(a)}{=}   \tau^{o}     \log _{2}\left(1+\frac{p}{p \sum_{i=1}^{K} \mathbb{E}\left\{\mathbf{a}_{k}^{H} \mathbb{E}\left\{\mathbf{e}_{i} \mathbf{e}_{i}^{H}\right\} \mathbf{a}_{k}\right\}+\sigma^{2} \mathbb{E}\{\left\|\mathbf{a}_{k}^{H}\right\|^{2}\}}\right)\\\label{lb_rate}
&\overset{(b)}{=}   \tau^{o}    \log _{2}\left(   1 + \frac{p}{ 
	 (  p\sum_{i=1}^{K}  \epsilon_i +\sigma^{2}) \;\mathbb{E}\{     
	[(\hat{\mathbf{Q}}^{H} \hat{\mathbf{Q}})^{-1}]_{k k}\}
}\right),
\end{align}
where $\epsilon_{i}$ is defined in (\ref{MSE_matrix}), $  (a) $ utilizes the independence between the channel estimate and the estimation errors, and $ (b)  $ is due to the result in (\ref{MSE_matrix}) and $  \left\|{\mathbf{a}}_{k}^{H}\right\|^{2}=\left[{\mathbf{A}}^{H} {\mathbf{A}}\right]_{k k}=[(\hat{\mathbf{Q}}^{H} \hat{\mathbf{Q}})^{-1}]_{k k}$.

\begin{theorem}\label{theorem2}
The achievable rate of user $k$ is lower bounded by
\begin{align}\label{rate}
&\underline{R_k}\left(\mathbf{\Phi}\right) =   \tau^{o}  \log _{2}\left( 1+   \frac{p\left(M-K\right)}{  \left(  p\sum_{i=1}^{K}  \epsilon_i +\sigma^{2}\right)  \left[\left(  \mathbf{\Lambda} +  \frac{\beta \delta}{\delta+1} \mathbf{H}_{1}^{H} \mathbf{\Phi}^{H} \mathbf{a}_{N} \mathbf{a}_{N}^{H} \mathbf{\Phi} \mathbf{H}_{1}   \right)^{-1}\right]_{k k}
	}\right),
\end{align}
where
$\mathbf{\Lambda} = \frac{\beta}{\delta+1} \mathbf{\Upsilon} \mathbf{H}_{1}^{H} \mathbf{H}_{1} \mathbf{\Upsilon}+\mathbf{\Omega} \boldsymbol{\Upsilon}^{2}+\frac{\sigma^{2}}{\tau p} \mathbf{\Upsilon}^{2}$.

\end{theorem}

\itshape {Proof:}  \upshape Please refer to Appendix \ref{App_3}. \hfill $\blacksquare$

The rate expression in Theorem \ref{theorem2} depends only on the slowly varying statistical CSI. Therefore, when designing the phase shifts to maximize the rate in (\ref{rate}), we only need to update the RIS's phase shifts over a much large time scale, which could effectively reduce overhead and computational complexity. Before the design of the phase shifts, we first analyze (\ref{rate}) to shed some light on the benefits of the RIS, and to answer the question whether RIS-aided massive MIMO is promising or not.


\subsection{Conventional Systems without RIS}
\begin{corollary}\label{corollary1}
When the RIS is switched off (i.e., $\alpha_{k}=\beta=0, \forall k$), the data rate (\ref{rate}) reduces to
\begin{align}\label{rate_without_RIS}
\begin{array}{l}
\underline{R_k}^{w/o} =  \tau^{o}  \log _{2}\left(1+\frac{p(M-K)}{p \sum_{i=1}^{K} \frac{1}{\frac{\tau p}{\sigma^{2}}+\frac{1}{ \gamma_{i}}}+\sigma^{2} }\times \frac{\gamma_{k}^{2}}{\gamma_{k}+\frac{\sigma^{2}}{\tau p}}\right).
\end{array}
\end{align}
\end{corollary}

When the RIS is switched off, the RIS-aided massive MIMO systems degrade to the conventional massive MIMO systems with Rayleigh fading channels ($\mathbf{Q}\to\mathbf{D}$), which has been studied in \cite{ngo2013energy}. As expected, the obtained rate (\ref{rate_without_RIS}) is the same as \cite[Eq. (42)]{ngo2013energy}. Based on (\ref{rate_without_RIS}), it can be seen that the rate is on the order of $\mathcal{O}\left(\log_2\left(M\right)\right)$, and the rate can maintain a non-zero value when the power is scaled down proportionally to $p=E_u/\sqrt{M}$, as the number of antennas $M\to\infty$, where $E_u$ is a constant. Specifically, we have
\begin{align}\label{power_scaling_conventional}
\lim\nolimits _{p=\frac{E_{u}}{\sqrt{M}}, M \rightarrow \infty} \;\;\underline{R_k}^{w/o} \rightarrow  \tau^{o}  \log _{2}\left(1+{\tau E_{u}^{2} \gamma_{k}^{2} {\sigma^{-4}} }\right).
\end{align}

Note that the achievable rate in (\ref{rate_without_RIS}) and power scaling law in (\ref{power_scaling_conventional}) will serve as baselines and help us identify the benefits enabled by introducing an RIS.

\subsection{What's New After Integrating An RIS?}
The order of magnitude of $\underline{R_{k}}\left(\mathbf{\Phi}\right)$ in (\ref{rate}) with respect to $M$ is $\mathcal{O}\left(\log_2\left(M\right)\right)$, since $\epsilon_k$ and $\mathbf{\Lambda}$ are independent of $M$. However, it is challenging to determine how $\underline{R_k}\left(\mathbf{\Phi}\right)$ scales with $N$, due to the unknown value of $\bf\Phi$ and the inverse operator.
For tractability, we propose an insightful lower bound $\underline{\underline{R_k}}$ for $\underline{R_k}\left(\mathbf{\Phi}\right)$ in the following.
\begin{corollary}\label{corollary2}
	A $\bf\Phi$-independent lower bound $\underline{\underline{R_{k}}}$ is given by
	\begin{align}\label{rate_much_lowBound}
& {\underline{R_{k}}} \left(\mathbf{\Phi}\right) \geq \underline{\underline{R_{k}}} =  \tau^{o}   \log _{2}\left( 1+   \frac{p\left(M-K\right)}{   \left(  p\sum_{i=1}^{K}  \epsilon_i +\sigma^{2}\right) \left[  \mathbf{\Lambda}^{-1}\right]_{k k} } 
\right),
\end{align}
	where equality holds when $\delta=0$, and the gap $\underline{R_{k}}\left(\mathbf{\Phi}\right)-\underline{\underline{R_{k}}}$ enlarges after optimizing $\bf\Phi$.
	Besides, (\ref{rate_much_lowBound}) can be approximated as
	\begin{align}\label{loweBound_approx}
	\underline{\underline{R_{k}}} &\approx  \tau^{o}    \log _{2}\left( 1+   \frac{p\left(M-K\right) }{   p\sum_{i=1}^{K}  \epsilon_i +\sigma^{2}} \times 
	\frac{\left(N \frac{\alpha_{k} \beta}{\delta+1}+\gamma_{k}\right)^{2}}{N \frac{\alpha_{k} \beta}{\delta+1}+\gamma_{k}+\frac{\sigma^{2}}{\tau p}} \right),
	\end{align}
	which scales on the order of $\mathcal{O}\left(\log_2\left(MN\right)\right)$. 
\end{corollary}

\itshape {Proof:}  \upshape Please refer to Appendix \ref{App_4}. \hfill $\blacksquare$


 Interestingly, if we treat $N \frac{\alpha_{k} \beta}{\delta+1}+\gamma_{k}$ as a new path-loss factor, (\ref{loweBound_approx}) possesses the same form as (\ref{rate_without_RIS}). This reveals two fundamental impacts of the RIS: $i)$ Positive effect: RIS enhances the channel strength by a factor $N \frac{\alpha_{k} \beta}{\delta+1}$; $ii)$ Negative effect: RIS results in larger channel estimation errors $\epsilon_k$. 
However, the channel strength always increases with $N$ since $\frac{(N \frac{\alpha_{k} \beta}{\delta+1}+\gamma_{k})^{2}}{N \frac{\alpha_{k} \beta}{\delta+1}+\gamma_{k}+\frac{\sigma^{2}}{\tau p}}$ is an increasing function of $N$, 
but the estimation error saturates to $\epsilon_k\to\frac{\sigma^2}{\tau p}$ as $N\to\infty$. Therefore, for large $N$, the benefits of the RIS outweigh its drawbacks in massive MIMO systems.


Corollary \ref{corollary2} proves that even with imperfect CSI, RIS-aided massive MIMO systems can achieve an ergodic rate at least on the order of $\mathcal{O}\left(\log_2\left(MN\right)\right)$. This promising gain comes from the additional $N$ paths contributed by the RIS for each user, such that more signals can be collected by the BS. 
Compared with $\mathcal{O}\left(\log_2\left(M\right)\right)$ in conventional systems, Corollary \ref{corollary2} proves that much higher capacity can be achieved after integrating an RIS. More importantly, the scaling law $\mathcal{O}\left(\log_2\left(MN\right)\right)$ indicates that if we want to maintain a fixed rate, the number of antennas can be reduced inversely proportional to the number of RIS elements. For better understanding, we provide a quantitative relationship for a special case.
\begin{corollary}\label{corollary3}
	When $\delta=0$ and for large $N$, to achieve $\mathrm{SNR}_k=\mathrm{C_0}$ for a given $N$, the required number of antennas $M$ is approximately given by
	\begin{align}\label{MN_tradeoff}
	M &\approx \frac{\mathrm{C_0}(K+\tau) \sigma^{2}}{\tau p\left(N \alpha_{k} \beta+\gamma_{k}\right)}+K
	=2\mathrm{C_0} \frac{\sigma^2}{p} \times \frac{1}{ N \alpha_{k} \beta+\gamma_{k}}+K, \text{  if  } \tau=K.
	\end{align}
\end{corollary}

\itshape {Proof:}  \upshape When $\delta=0$, we have $ {\underline{R_{k}}} \left(\mathbf{\Phi}\right)= \underline{\underline{R_{k}}}  $. Then, using (\ref{loweBound_approx}), for large $N$, we have $\epsilon_k\approx \frac{\sigma^2}{\tau p}$, and $ \mathrm{S N R}_{k} \approx \frac{p(M-K)}{K \frac{\sigma^{2}}{\tau}+\sigma^{2}}\left(N \alpha_{k} \beta+\gamma_{k}\right)$. Solving the equation $\mathrm{S N R}_{k}=\mathrm{C_0}$ completes the proof. \hfill $\blacksquare$

Corollary \ref{corollary3} corresponds to the scenarios with rich scattering.
Eq. (\ref{MN_tradeoff}) clearly exhibits the inverse proportional relationship between $M$ and $N$. Meanwhile, intuitively, $M$ increases with $\mathrm{C_0}$, $K$, and $\frac{\sigma^2}{p}$, but decreases with the link strengths $ \alpha_{k} \beta$ and $\gamma_k$. Since the RIS's reflecting elements consume much less energy than RF chains, Corollary \ref{corollary3} states that the energy efficiency can be remarkably improved by integrating an RIS.

\begin{corollary}\label{corollary4}
If the RIS-BS channel is purely LoS ($\delta\to\infty$), RIS-aided massive MIMO systems perform no worse than conventional massive MIMO systems, i.e., $\underline{R_k} \left(\mathbf{\Phi}\right)\geq \underline{R_k}^{w/o}$.
\end{corollary}

\itshape {Proof:} \upshape Substituting $\delta\to\infty$ into (\ref{rate_much_lowBound}), $\epsilon_k$, and $\kappa_k$, it can be shown that $\underline{\underline{R_k}} = \underline{R_k}^{w/o}$. Then, we have $\underline{R_k} \left(\mathbf{\Phi}\right)\geq \underline{\underline{R_{k}}} = \underline{R_k}^{w/o}$.  \hfill $\blacksquare$

Corollary \ref{corollary4} corresponds to the scenario where the RIS is carefully deployed to reduce the scatters and obstacles between the BS and the RIS. In this case, the additional channel estimation error in $\epsilon_k,\forall k$, caused by the RIS, vanishes. Therefore, the RIS only has the positive effect of enhancing the channel strength, which improves the achievable rate.
We emphasize that even though we can only prove that RIS-aided systems are no worse than conventional systems when $\delta\to\infty$, in general, it could perform much better because the second lower bound $\underline{\underline{R_{k}}} $ is not as tight as the first lower bound $\underline{R_k}\left(\mathbf{\Phi}\right)$ if $\bf\Phi$ is carefully designed.

%
%

\subsection{Power Scaling Law}
In conventional massive MIMO systems, an attractive feature is that the transmit power can be scaled down proportionally by increasing $M$ \cite{zhang2014power,ngo2013energy,bjornson2017massive}. After introducing an RIS, we reveal a new power scaling law with respect to $N$, and compare it to (\ref{power_scaling_conventional}).
\begin{corollary}\label{corollary5}
As $N\to\infty$, when the power is scaled proportionally to $p=E_u/N$, the achievable rate in (\ref{rate}) can maintain a non-zero value $ \underline{\vec{R}_{k}}\left(\mathbf{\Phi}\right) \rightarrow \tau^{o} \log _{2}\left(1+\overrightarrow{\mathrm{SNR}}_{k}\right) $, where
			\begin{align}\label{power_LB}
			\overrightarrow{\mathrm{SNR}}_{k} &= 			\frac{E_{u}(M-K)}{\sum_{i=1}^{K} \frac{E_{u}}{\frac{\tau E_{u}}{\sigma^{2}}+\frac{\delta+1}{\alpha_{i} \beta}}+\sigma^{2}}\times  \frac{1}{\left[  \mathbf{\Xi}^{-1}\right]_{kk}} \geq \frac{E_{u}(M-K)}{\sum_{i=1}^{K} \frac{E_{u}}{\frac{\tau E_{u}}{\sigma^{2}}+\frac{\delta+1}{{\alpha_{i} \beta}}}+\sigma^{2}} \times \frac{\left(\frac{\alpha_{k} \beta}{\delta+1}\right)^{2}}{\frac{\alpha_{k} \beta}{\delta+1}+\frac{\sigma^{2}}{\tau E_{u}}}.
		\end{align}
		with
		$\begin{array}{l}
		\mathbf{\Xi}=\operatorname{diag}\left\{\frac{(\frac{\alpha_{1} \beta}{\delta+1})^{2}}{\frac{\alpha_{1} \beta}{\delta+1}+\frac{\sigma^{2}}{\tau E_{u}}}, \ldots, \frac{(\frac{\alpha_{K} \beta}{\delta+1})^{2}}{\frac{\alpha_{K} \beta}{\delta+1}+\frac{\sigma^{2}}{\tau E_{u}}}\right\}+\frac{\beta \delta}{\delta+1} \frac{\mathbf{H}_{1}^{H} \boldsymbol{\Phi}^{H} \mathbf{a}_{N} \mathbf{a}_{N}^{H} \boldsymbol{\Phi} \mathbf{H}_{1}}{N}.
		\end{array}$
\end{corollary}

\itshape {Proof:} \upshape 
Substitute $p=\frac{E_{u}}{N}$ into (\ref{rate}). As $N\to\infty$, we have $\kappa_{k} \rightarrow \frac{\frac{\alpha_{k} \beta}{\delta+1}}{\frac{\alpha_{k} \beta}{\delta+1}+\frac{\sigma^{2}}{\tau E_{u}}}$, $\frac{E_u}{N} \epsilon_{i}\rightarrow \frac{E_{u}}{\frac{\delta+1}{\alpha_{i} \beta}+\frac{\tau E_{u}}{\sigma^{2}}}$,  $ \frac{\mathbf{\Upsilon} \mathbf{H}_{1}^{H} \mathbf{H}_{1} \mathbf{\Upsilon}}{N} \rightarrow \operatorname{diag}\left\{\kappa_{1}^{2} \alpha_{1}, \ldots, \kappa_{K}^{2} \alpha_{K}\right\} $, $ \frac{\mathbf{\Omega} \mathbf{\Upsilon}^{2}}{N} \rightarrow \mathbf{0} $, and
$ \frac{\sigma^{2}}{ \tau p} \frac{\mathbf{\Upsilon}^{2}}{N} \rightarrow \frac{\sigma^{2}}{\tau E_{u}} \mathbf{\Upsilon}^{2} $, which help us arrive at the first equation in (\ref{power_LB}). Then, using the inequality in (\ref{inverse_lowerBound}), we can obtain the lower bound.
\hfill $\blacksquare$

Comparing (\ref{power_LB}) with (\ref{power_scaling_conventional}), it can be seen that this new scaling law has a high order of magnitude with respect to $M$. Besides, by comparing (\ref{power_LB}) with (\ref{rate_without_RIS}), it is interesting to find that (\ref{power_LB}) can be interpreted as the SNR achieved by a conventional massive MIMO system with transmit power $E_u$ and path-loss $\frac{\alpha_{k} \beta}{\delta+1}$.
To sum up, for large $M$ and $N$, transmit power can be significantly reduced while achieving high data rates.

\subsection{Comparison with MRC-based Systems}
\begin{corollary}\label{corollary7}
	When $p$ or $M$ or $N$ is large, ZF-based RIS-aided massive MIMO outperforms its MRC-based counterpart. Besides, the severe fairness problem in MRC-based RIS-aided massive MIMO system \cite[Remark 2]{zhi2021twotimescale} does not exist in the considered ZF-based systems.
\end{corollary}

\itshape {Proof:} \upshape According to Corollary \ref{corollary2}, when $p$ or $M$ grows without bound, it is found that $R_k\geq \underline{\underline{R_{k}}}\to\infty,\forall k$. Thus, all users can have infinite data rates.  However, as proved in \cite[Remark 2]{zhi2021twotimescale}, when using MRC detectors, due to the mutual interference, the rate is still bounded when $p$ or $M$ is large. 
Meanwhile, the rates of all users in the considered system are at least on the order of $\mathcal{O}\left(\log_2\left(N\right)\right)$. However, when using MRC, the rate of only one user can be on the order of $\mathcal{O}\left(\log_2\left(N\right)\right)$, while the rates of all other users degrade to zero when $N$ is large, which results in a serious fairness problem.
\hfill $\blacksquare$

ZF-based RIS systems perform better since RIS-aided systems suffer from severe multi-user interference.
This is because multiple users share the common RIS-BS channel, and thus the $K$ users' channels are highly correlated. The highly correlated channels result in severe interference and low data rate. 
However, by using ZF, the severe multi-user interference issue can be addressed, which leads to promising performance for various aspects.

\subsection{The Upper Bound}
The analysis based on the lower bound $\underline{\underline{R_{k}}}$ is rigorous but conservative, since it ignores the performance gain achieved by optimizing $\bf\Phi$. 
We next provide an upper bound to unveil the maximum gain achieved by optimizing $\bf\Phi$.
\begin{corollary}\label{corollary9}
	The rate is upper bounded by $ \underline{R_{k}}(\boldsymbol{\Phi}) \leq \overline{R_{k}} = \tau^{o} \log _{2}\left(1+\overline{\mathrm{SNR}}_{k}\right) $, where
	\begin{align}\label{up_general}
	\overline{\mathrm{SNR}}_{k}&=
	\frac{p(M-K)}{p \sum_{i=1}^{K} \epsilon_{i}+\sigma^{2}}\left\{
	\frac{(N \frac{\alpha_{k} \beta}{\delta+1}+\gamma_{k})^{2}}{N \frac{\alpha_{k} \beta}{\delta+1}+\gamma_{k}+\frac{\sigma^{2}}{\tau p}} +     \left|\mathbf{a}_{N}^{H} \mathbf{\Phi} \overline{\mathbf{h}}_{k}\right|^{2}\frac{\alpha_{k} \beta \delta}{\delta+1}
	\right\}\\\label{rate_upBound}
	&\leq \frac{p(M-K)}{p \sum_{i=1}^{K} \epsilon_{i}+\sigma^{2}}\left\{
	\frac{(N \frac{\alpha_{k} \beta}{\delta+1}+\gamma_{k})^{2}}{N \frac{\alpha_{k} \beta}{\delta+1}+\gamma_{k}+\frac{\sigma^{2}}{\tau p}} +N^{2} \frac{\alpha_{k} \beta \delta}{\delta+1} 
	\right\}.
	\end{align}	
	
	Based on (\ref{up_general}), $\overline{R_{k}}$ is at least on the order of $\mathcal{O}\left(\log_2\left(MN\right)\right)$. Based on (\ref{rate_upBound}), $\overline{R_{k}}$ is on the order of $\mathcal{O}\left(\log_2\left(MN^2\right)\right)$. 
\end{corollary}

\itshape {Proof:} \upshape Please refer to Appendix \ref{App_8}. \hfill $\blacksquare$

We emphasize that (\ref{up_general}) holds for all $K$ users but (\ref{rate_upBound}) does not. This is because (\ref{rate_upBound}) is achieved by aligning the RIS phase shifts to a specific user $k$, i.e., $\mathbf{a}_{N}^{H} \boldsymbol{\Phi} \overline{\mathbf{h}}_{k} = N$. However, when $\mathbf{a}_{N}^{H} \boldsymbol{\Phi} \overline{\mathbf{h}}_{k} = N$, it is known that $\mathbf{a}_{N}^{H} \boldsymbol{\Phi} \overline{\mathbf{h}}_{i}$, $\forall i\neq k$, is bounded even for $N\to\infty$\cite{zhi2021twotimescale}.
Thus, the additional $N$-fold gain in (\ref{rate_upBound}) comes from the concentration of passive beamforming on user $k$.
 Combining the lower bound in Corollary \ref{corollary2} and this upper bound, we highlight the following conclusion: 
 \begin{remark}\label{remark3}
 If we align the RIS phase shifts for one user, the rate of this user will scale \textbf{at most} on the order of $\mathcal{O}\left(\log_2\left(MN^2\right)\right)$, while the rates of the other users scale \textbf{at least} on the order of $\mathcal{O}\left(\log_2\left(MN\right)\right)$, which is high as well. 
 \end{remark}
 
 Based on these two achievable rate scaling laws, the sum user rate will be high for large $M$ and $N$, if we simply align the RIS phase shifts for an arbitrary user, which constitutes a low-complexity heuristic approach for the sum-rate maximization problem.

\begin{corollary}\label{corollary10}
	The quantization error caused by RIS discrete phase shifts does not impact the derived achievable rate scaling orders.
\end{corollary}

\itshape {Proof:} \upshape First, the lower bound $\underline{\underline{R_{k}}}$ does not depend on $\bf\Phi$, and hence, is not affected by quantization errors. Secondly, $\left|\mathbf{a}_{N}^{H} \mathbf{\Phi} \overline{\mathbf{h}}_{k}\right|^{2}\geq N^{2} \cos ^{2}\left(\frac{\pi}{2^{b}}\right)$ holds for an RIS with $b$-bit quantization\cite{han2019large}. Therefore, scaling order $\mathcal{O}\left(\log_2\left(MN^2\right)\right)$ still holds for $\overline{R_{k}}$.  \hfill $\blacksquare$

\subsection{Summary}
We summarize that RIS-aided massive MIMO with ZF detectors is promising for 
\begin{itemize}
\item \textbf{Green communications (Corollary \ref{corollary3}) }: The number of BS antennas can be reduced inversely proportional to the number of RIS elements, while maintaining a constant rate.
\item \textbf{Enhanced mobile broadband (Corollary \ref{corollary2}, \ref{corollary9}, \ref{corollary10}, Remark \ref{remark3})} : According to the rate scaling orders, ultra-high throughput requirement can be achieved for large $M$ and $N$.
\item \textbf{Internet of things (Corollary \ref{corollary5})} : For large $M$ and $N$, all users can significantly reduce their transmit powers while maintaining high data rates.
\end{itemize}

\section{RIS Phase Shift Design}\label{section4}
In this section, based on the derived rate expression in (\ref{rate}) and the low-complexity MM technique\cite{sun2017MM}, we aim to solve the sum user rate maximization (Max-Sum) and the minimum user rate maximization (Max-Min) problems, respectively. 
The Max-Sum problem maximizes the utility but may sacrifice fairness. On the contrary, the Max-Min problem guarantees fairness but may sacrifice utility. Thus, simultaneously investigating both problems can help us understand which optimization criterion is more suitable for the considered systems. For tractability, variable $\bf\Phi$ is rewritten as $\mathbf{\Phi}=\operatorname{diag}\left\{\mathbf{v}^{H}\right\}$, where $\mathbf{v}=\left[e^{j \theta_{1}}, \ldots, e^{j \theta_{N}}\right]^{H}$. Then, we can transform the design of $\bf\Phi$ to the design of vector $\bf v$.

\begin{lem}
The rate in (\ref{rate}) can be rewritten as $ \underline{R_k} \left(\mathbf{v}\right)= \frac{ \tau^{o}  }{  \ln(2)} \ln \left(1+\frac{\mathbf{v}^{H} \mathbf{B} \mathbf{v}}{\mathbf{v}^{H} \mathbf{C}_{k} \mathbf{v}}\right) $,
where
\begin{align}\label{BCk}
\begin{array}{l}
\mathbf{B}=
\frac{1}{N} \mathbf{I}_{N}+\frac{\beta \delta}{\delta+1} \operatorname{diag}\left\{\mathbf{a}_{N}^{H}\right\}  \mathbf{H}_{1} \mathbf{\Lambda}^{-1} \mathbf{H}_{1}^{H} \operatorname{diag}\left\{\mathbf{a}_{N}\right\}, \\
\mathbf{C}_{k}=
\frac{     p\sum_{i=1}^{K}  \epsilon_i +\sigma^{2}  }{p(M-K)}
\left(\left[\mathbf{\Lambda}^{-1}\right]_{k k} \mathbf{B}-\frac{\beta \delta}{\delta+1} \mathbf{z}_{k} \mathbf{z}_{k}^{H}\right),
\end{array}
\end{align}
and $ \mathbf{z}_{k}^{H}=\left[\boldsymbol{\Lambda}^{-1} \mathbf{H}_{1}^{H} \operatorname{diag}\left\{\mathbf{a}_{N}\right\}\right]_{(k, :)} $. Besides, we have ${\bf B}\succ {\bf 0}$ and $\mathbf{C}_k\succeq \mathbf{0}$.
\end{lem}

\itshape {Proof:}  \upshape We can complete the proof by substituting the last equality in (\ref{inverse_lowerBound}) into (\ref{rate}), and using $\mathbf{\Phi}^{H} \mathbf{a}_{N}=\operatorname{diag}\left\{\mathbf{a}_{N}\right\} \mathbf{v}$ and $1=\frac{1}{N}\mathbf{v}^H\mathbf{I}_N\mathbf{v}$. Besides, we have $\mathbf{B}\succ \mathbf{0}$ due to $\mathbf{\Lambda}^{-1}\succ \mathbf{0}$, which results in $\mathbf{v}^{H} \mathbf{B} \mathbf{v}>0$.
 Since the rate $ \underline{R_k} \left(\mathbf{v}\right)$ must be non-negative due to its definition in (\ref{positive_rate}), we obtain $\mathbf{v}^{H} \mathbf{C}_k \mathbf{v}\geq0$, which means that $\mathbf{C}_k\succeq \mathbf{0}$.  \hfill $\blacksquare$

Define $ f_{k}(\mathbf{v}) \triangleq \ln \left(1+\frac{\mathbf{v}^{H} \mathbf{B v}}{\mathbf{v}^{H} \mathbf{C}_{k} \mathbf{v}}\right) $ for brevity. Since the same factor $\frac{ \tau^{o} }{ \ln(2)} $ is included in $\underline{R_k} \left(\mathbf{v}\right),\forall k$, we can ignore it and formulate the following two optimization problems
\begin{align}\label{MaxSum}
&\textbf{Max-Sum}:  \max_{\mathbf{v}}  \; \sum\nolimits_{k=1}^{K} f_{k}(\mathbf{v}), \qquad\quad { s.t. } \; \left|[\mathbf{v}]_{(n)}\right|=1, \forall n. \\\label{MaxMin}
 & \textbf{Max-Min}:   \max_{\mathbf{v}} \;\; \min_{k} \;\;  f_{k}(\mathbf{v}), \;\qquad\;\quad { s.t. } \; \left|[\mathbf{v}]_{(n)}\right|=1, \forall n. 
\end{align}

To successfully solve the above two problems under the MM algorithm framework, tractable lower-bound surrogate functions need to be constructed for objective functions in (\ref{MaxSum}) and (\ref{MaxMin}), and then closed-form optimal solutions are expected to be derived via the surrogate functions. 

\subsection{Max-Sum Problem}
\begin{lem}\label{lem_surrogate}
For a fixed point $\mathbf{v}_n$, a lower bound of ${f_k} (\mathbf{v})  $ is given by
\begin{align}\label{fk_lb}
f_{k}(\mathbf{v})  \geq \underline{f_k} (\mathbf{v} \mid \mathbf{v}_n) = \operatorname {const} _{k}+2 \operatorname{Re}\left\{\left(\mathbf{f}_{k}^{n}\right)^H \mathbf{v}\right\},
\end{align}
where 
\begin{align}\label{omiga_psi}
&\operatorname {const} _{k}=f_{k}\left(\mathbf{v}_{n}\right)-\frac{   \mathbf{v}_{n}^H   \mathbf{B v}_{n}}{   \mathbf{v}_{n}^H    \mathbf{C}_{k} \mathbf{v}_{n}}-\psi_{k}   \mathbf{v}_{n}^H   \left(\lambda_{\max }\left(\mathbf{C}_{k}+\mathbf{B}\right) \mathbf{I}_N-\left(\mathbf{C}_{k}+\mathbf{B}\right)\right) \mathbf{v}_{n}-N \psi_{k} \lambda_{\max }\left(\mathbf{C}_{k}+\mathbf{B}\right),\nonumber\\
&\left(\mathbf{f}_{k}^{n}\right)^H=\omega_{k}   \mathbf{v}_{n}^H   \mathbf{B}-\psi_{k} \mathbf{v}_{n}^H  \left(\left(\mathbf{C}_{k}+\mathbf{B}\right)-\lambda_{\max }\left(\mathbf{C}_{k}+\mathbf{B}\right) \mathbf{I}_N\right),\nonumber\\
&\omega_{k} =\frac{1}{\mathbf{v}_{n}^H \mathbf{C}_{k} \mathbf{v}_{n}}, \qquad \psi_{k} =\frac{ \mathbf{v}_{n}^H  \mathbf{B} \mathbf{v}_{n}}{\left(  \mathbf{v}_{n}^H   \mathbf{C}_{k} \mathbf{v}_{n}\right)\left( \mathbf{v}_{n}^H  \mathbf{C}_{k} \mathbf{v}_{n}+  \mathbf{v}_{n}^H  \mathbf{B v}_{n}\right)}.
\end{align} 
\end{lem}

\itshape {Proof:}  \upshape Please refer to Appendix \ref{App_6}. \hfill $\blacksquare$

Then, the Max-Sum problem (\ref{MaxSum}) can be directly solved based on the proposed surrogate function $\underline{f_k} (\mathbf{v} \mid \mathbf{v}_n)$ in Lemma \ref{lem_surrogate}. Denoted by $\mathbf{v}_n$ the solution in the $n$-th iteration, the closed-form optimal solution in the $\left(n+1\right)$-th iteration is given by
\begin{align}\label{solution_MaxSum}
\mathbf{v}_{n+1}
=\arg \max_{\mathbf{v}}  \sum\nolimits_{k=1}^{K} \underline{{f}_{k}}\left(\mathbf{v} \mid \mathbf{v}_n \right)
=\exp \left\{j \angle\left(\sum\nolimits_{k=1}^{K} \mathbf{f}_{k}^{n}\right)\right\}.
\end{align}

\subsection{Max-Min Problem}
Next,  we focus on the Max-Min problem (\ref{MaxMin}), which is more challenging since the objective function $\min\limits_{k}  f_{k}(\mathbf{v})$ is non-differential. Therefore, we first adopt the log-sum-exp approximation in \cite{xingsi1992entropy} to obtain a lower-bounded smooth objective function, as follows
\begin{align}\label{tilde_f}
\begin{aligned}
&\min\limits_{k}  f_{k}(\mathbf{v}) \geq \min\limits_{k}   \underline{f_k} (\mathbf{v} \mid \mathbf{v}_n)   \geq \widetilde{f}\left(\mathbf{v}\right) \triangleq -\frac{1}{\mu} \ln \left(\sum\nolimits_{k=1}^K \exp \left\{-\mu  \underline{f_k} (\mathbf{v} \mid \mathbf{v}_n)    \right\}\right)     ,
\end{aligned}
\end{align}
where $\mu>0$ is a constant for controlling the approximation accuracy, and the last inequality can be proved similar as \cite[(15)]{xingsi1992entropy}.
\begin{lem}
	For a fixed point $\mathbf{v}_n$, $\widetilde{f}\left(\mathbf{v}\right)$ in (\ref{tilde_f}) is lower bounded by
\begin{align}
\begin{aligned}
 \widetilde{f}\left(\mathbf{v}\right)\geq  \underline{\widetilde{f}}\left(\mathbf{v} \mid \mathbf{v}_n \right)&= \widetilde{\mathrm{const}} +2 \operatorname{Re}\left\{\left[\left(\sum\nolimits_{k=1}^K l_{k}^{n}\left(\mathbf{f}_{k}^{n}\right)^{H}\right)+\left(2 \mu \max _{k}\left\|\mathbf{f}_{k}^{n}\right\|^{2}\right)\mathbf{v}_{n}^{H}\right] \mathbf{v}\right\},
\end{aligned}
\end{align}
where
\begin{align}
& \widetilde{\mathrm{const}} = \widetilde{f}\left(\mathbf{v}_{n}\right)-2 \operatorname{Re}\left\{\sum\nolimits_{k=1}^K l_{k}^{n}\left(\mathbf{f}_{k}^{n}\right)^{H} \mathbf{v}_{n}\right\}+2 N\left(-2 \mu \max _{k}\left\|\mathbf{f}_{k}^{n}\right\|^{2}\right),\\\label{lkn}
&	l_{k}^{n}=\frac{\exp \left\{-\mu  \underline{f_k} (\mathbf{v}_n \mid \mathbf{v}_n)  \right\}}{\sum_{k=1}^K \exp \left\{-\mu \underline{f_k} (\mathbf{v}_n \mid \mathbf{v}_n) \right\}}.
\end{align}
\end{lem}

\itshape {Proof:}  \upshape Please refer to Appendix \ref{App_7}. \hfill $\blacksquare$

Based on the MM algorithm, the Max-Min problem (\ref{MaxMin}) can be solved by maximizing the lower bound $ \underline{\widetilde{f}}\left(\mathbf{v} \mid \mathbf{v}_n \right)$ in each iteration. Given the solution $\mathbf{v}_n$ in the $n$-th iteration, the closed-form optimal solution at the $(n+1)$-th iteration is
\begin{align}\label{solution_MaxMin}
\mathbf{v}_{n+1}
=\arg \max_{\mathbf{v}} \underline{\widetilde{f}}\left(\mathbf{v} \mid \mathbf{v}_n \right) =\exp \left\{j\angle \left\{\left(\sum\nolimits_{k=1}^K l_{k}^{n}\mathbf{f}_{k}^{n}\right)+\left(2 \mu \max _{k}\left\|\mathbf{f}_{k}^{n}\right\|^{2}\right)\mathbf{v}_{n}\right\}\right\}.
\end{align}

Finally, the framework for solving Max-Sum problem (\ref{MaxSum}) and Max-Min problem (\ref{MaxMin}) are summarized in Algorithm \ref{algorithm1}, where steps $4-9$ are used to accelerate the convergence of the MM technique\cite{zhou2020multicast}.

\begin{algorithm}
	\caption{MM algorithm.}
	\begin{algorithmic}[1]\label{algorithm1}
		\STATE Initialize $\mathbf{v}_0$, $n=0$;
		\REPEAT 
		\STATE Given $\mathbf{v}_n$, obtain solution $\mathbf{v}_{n+1}^{(1)}$ from (\ref{solution_MaxSum}) or (\ref{solution_MaxMin});
		\STATE Given $\mathbf{v}_{n+1}^{(1)}$, obtain solution $\mathbf{v}_{n+1}^{(2)}$ from (\ref{solution_MaxSum}) or (\ref{solution_MaxMin});
		\STATE $\triangle_{\mathbf{v}1} = \mathbf{v}_{n+1}^{(1)} - \mathbf{v}_n$, and $\triangle_{\mathbf{v}2} = \mathbf{v}_{n+1}^{(2)} - \mathbf{v}_{n+1}^{(1)} - \triangle_{\mathbf{v}1}$;
		\STATE $ \rho=-\frac{\left\|     \triangle_{\mathbf{v}1}   \right\|}{\left\|   \triangle_{\mathbf{v}2}   \right\|} $, and $\mathbf{v}_{n+1} = -\exp \left\{j\angle\left( \mathbf{v}_n - 2\rho \triangle_{\mathbf{v}1} +\rho^2 \triangle_{\mathbf{v}2}   \right)\right\}$;
		\WHILE {$\mathbf{v}_{n+1} $ does not lead to an increasing objective value in (\ref{MaxSum}) or (\ref{MaxMin}) }
		\STATE $\rho=\left(\rho-1\right)/2$, and $\mathbf{v}_{n+1} = -\exp \left\{j\angle\left( \mathbf{v}_n - 2\rho \triangle_{\mathbf{v}1} +\rho^2 \triangle_{\mathbf{v}2}  \right)\right\}$;
		\ENDWHILE
		\STATE $ n\leftarrow n+1 $;
		\UNTIL The objective value in (\ref{MaxSum}) or (\ref{MaxMin}) converges.
	\end{algorithmic}
\end{algorithm}

\section{Numerical Results}\label{section5}
In this section, we verify the correctness of our derived results and give insights.
Unless otherwise stated, as in \cite{zhi2021twotimescale}, we set $K=8$, $M=N=64$, $\delta=1$, $\tau_c=196$, $\tau=K$, $p=30$ dBm, $\sigma^2=-104$ dBm and $ \mu=10 $. 
The BS and the RIS are located at $(0,0)$ and $(0,700)$, respectively. The users are randomly located at a circle centred at $(10,700)$ of radius $10$ m. The path-loss, the AoA and AoD are set the same values in \cite{zhi2021twotimescale}. The theoretical result in (\ref{rate}) is verified via Monte-Carlo simulations based on (\ref{exact_rate}). The MRC-based system for perfect and imperfect CSI are evaluated based on \cite{zhi2020directLinks} and \cite{zhi2021twotimescale}, respectively.

To begin with, we evaluate the lower bounds $\underline{\underline{R_{k}}}$ in (\ref{rate_much_lowBound}) and (\ref{loweBound_approx}), and the upper bound $\overline{R_{k}}$ in (\ref{up_general}) and (\ref{rate_upBound}), respectively. 
Without loss of generality, we denote the user nearest to and furthest from the RIS as users $1$ and $8$, respectively. Four phase shifts designs are considered for the RIS:
\begin{itemize}
\item \textbf {Case 1:} Align the phase shifts to the nearest user $1$, i.e., $\mathbf{a}_{N}^{H} \mathbf{\Phi} \overline{\mathbf{h}}_{1} = N$. 
\item \textbf {Case 2:} Align the phase shifts to the furthest user $8$, i.e., $\mathbf{a}_{N}^{H} \mathbf{\Phi} \overline{\mathbf{h}}_{8} = N$. 
\item \textbf {Case 3:} Set the phase shifts $\theta_{n},\forall n$, randomly in $\left[0,2\pi\right]$. 
\item \textbf {Case 4:} Set $\mathbf{\Phi} = \mathbf{I}_N$. 
\end{itemize}

\begin{figure}[htbp]
	\centering    
	
	\subfigure[Rate of user 1 or user 8]  
	{
		\begin{minipage}{8cm}
			\centering          
			\includegraphics[scale=0.4]{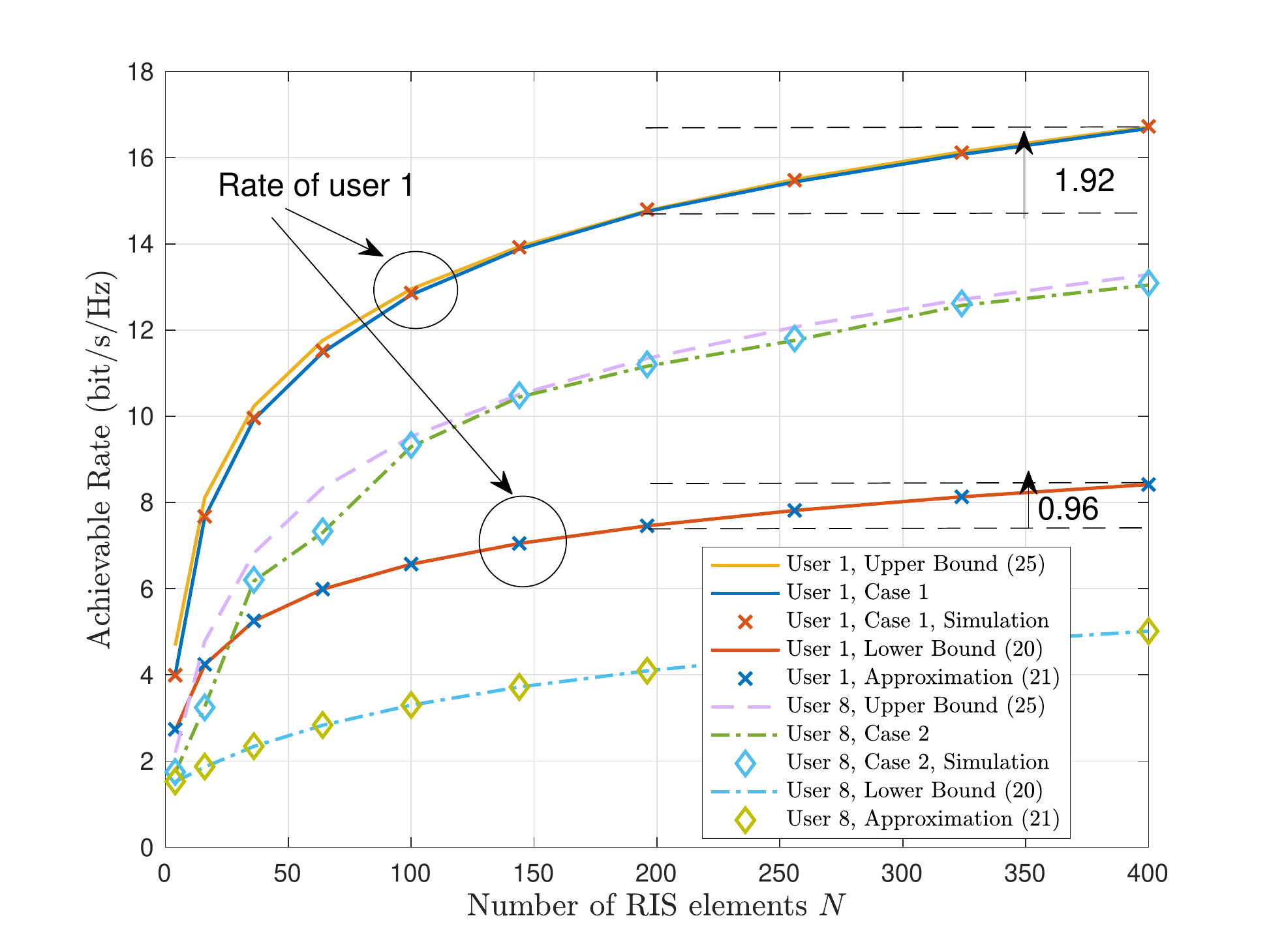}   
		\end{minipage}\label{figure2a}
	}
	\subfigure[Rate of user 1 in Case 2 - 4] 
	{
		\begin{minipage}{7cm}
			\centering      
			\includegraphics[scale=0.4]{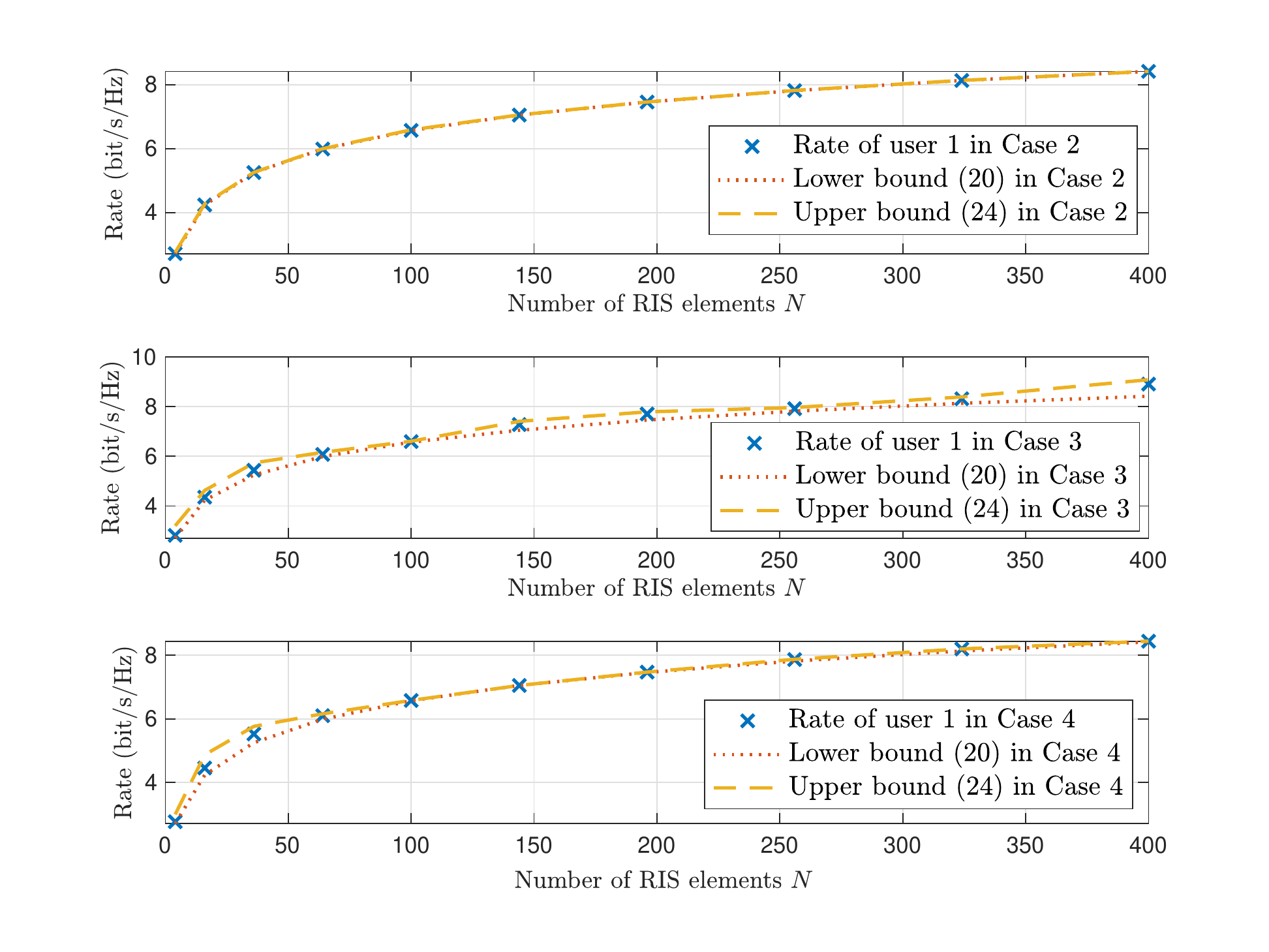}   
		\end{minipage}\label{figure2b}
	}
	
	\caption{Rate of a user under different RIS designs.} 
	\label{figure2}  
\end{figure}
Fig. \ref{figure2a} illustrates the rate of one user when the RIS phase shifts are aligned to it. To be specific, we respectively plot the rate of user 1 in Case 1, and the rate of user 8 in Case 2. We also plot the corresponding upper and lower bounds.
Firstly, we observe that when the RIS phase shifts are aligned to user 1 or user 8, their rates tightly approach the upper bound in (\ref{rate_upBound}),  which validates that the derived scaling order $\mathcal{O}\left(\log_2\left(MN^2\right)\right)$ in (\ref{rate_upBound}) is achievable.  Secondly, the theoretical results match well with the simulation results, which verifies the correctness of our derivatives. Besides, it is seen that user 1 has better performance than user 8, since it locates closer to the RIS and then has a higher path-loss factor.
Thirdly, we can see that the approximate lower bound (\ref{loweBound_approx}) perfectly matches with the accurate lower bound (\ref{rate_much_lowBound}) for all considered values of $N$, which verifies the reliability of our previous analysis based on (\ref{rate_much_lowBound}). Finally, when $N$ is doubled from $N=200$ to $N=400$, the increment of the rate in lower bound and that in Case 1 are almost $\tau^o\log_{2}\left(2\right)=0.96$ and $\tau^o\log_{2}\left(2^2\right)=1.92$, respectively, which confirms the theoretical scaling orders $\mathcal{O}\left(\log_2\left(MN\right)\right)$ and $\mathcal{O}\left(\log_2\left(MN^2\right)\right)$. 

Fig. \ref{figure2b} shows the rate of user 1 when the RIS phase shifts
\itshape {are not}  \upshape
aligned to it, i.e., in Case 2 - 4. We can observe that in these three cases, the upper bound (\ref{up_general}) and lower bound (\ref{rate_much_lowBound}) are tight, which means that the rate scales accurately on the order of $\mathcal{O}\left(\log_2\left(MN\right)\right)$.
 This is because the RIS phase shifts cannot be aligned simultaneously to many users. Then, only one user's rate can scale as $\mathcal{O}\left(\log_2\left(MN^2\right)\right)$ while the rates of all other users scale only as $\mathcal{O}\left(\log_2\left(MN\right)\right)$.
Therefore, the scaling order $\mathcal{O}\left(\log_2\left(MN\right)\right)$ obtained based on the lower bound is appropriate for understanding the system capacity since it corresponds to the rate of most of the users. 

\begin{figure}[htbp]
	\centering   
	
	\subfigure[Sum user rate] 
	{
		\begin{minipage}{7cm}
			\centering          
			\includegraphics[scale=0.4]{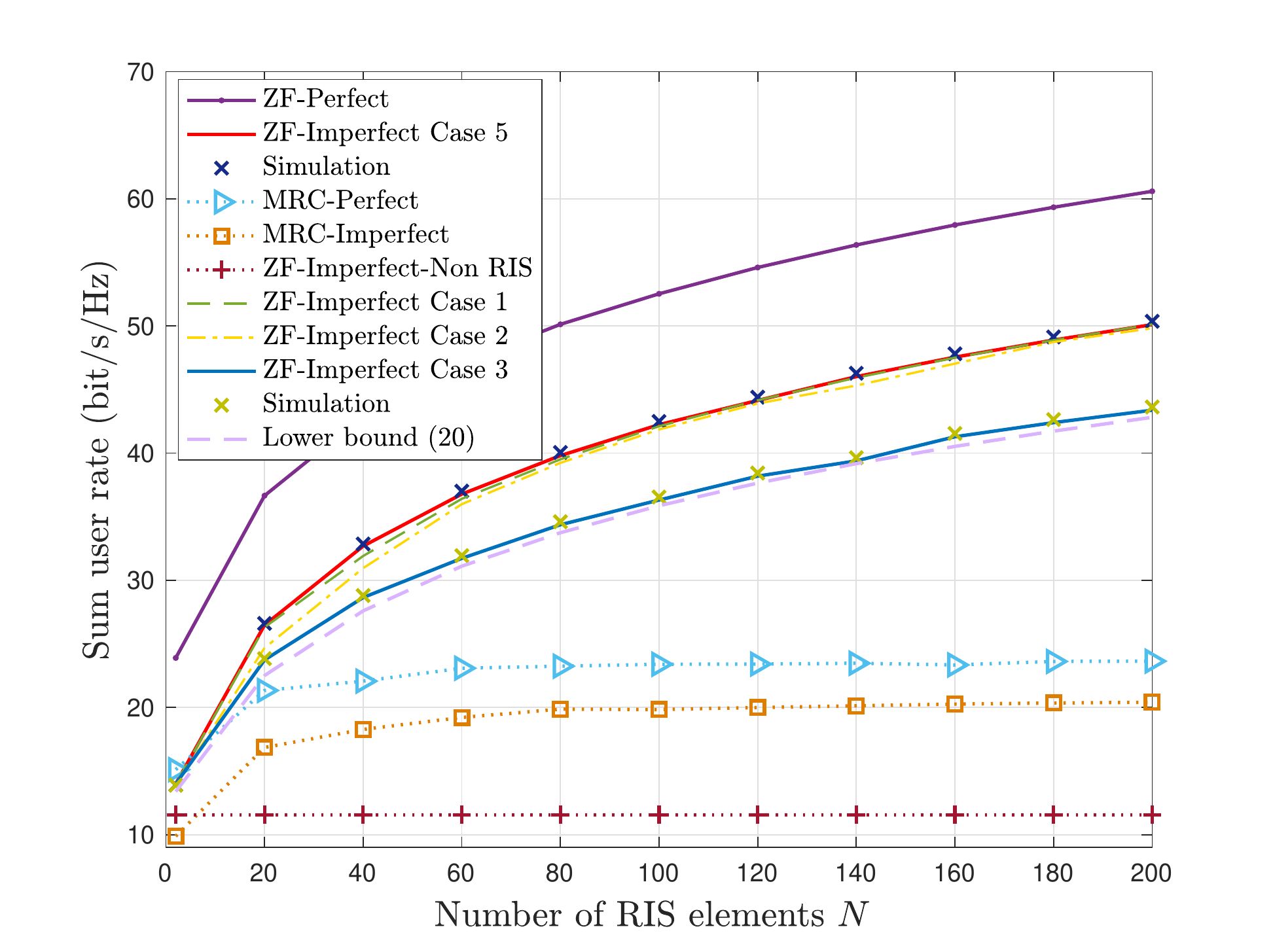}   
		\end{minipage}\label{figure3a}
	}
	\subfigure[Minimum user rate] 
	{
		\begin{minipage}{8cm}
			\centering      
			\includegraphics[scale=0.4]{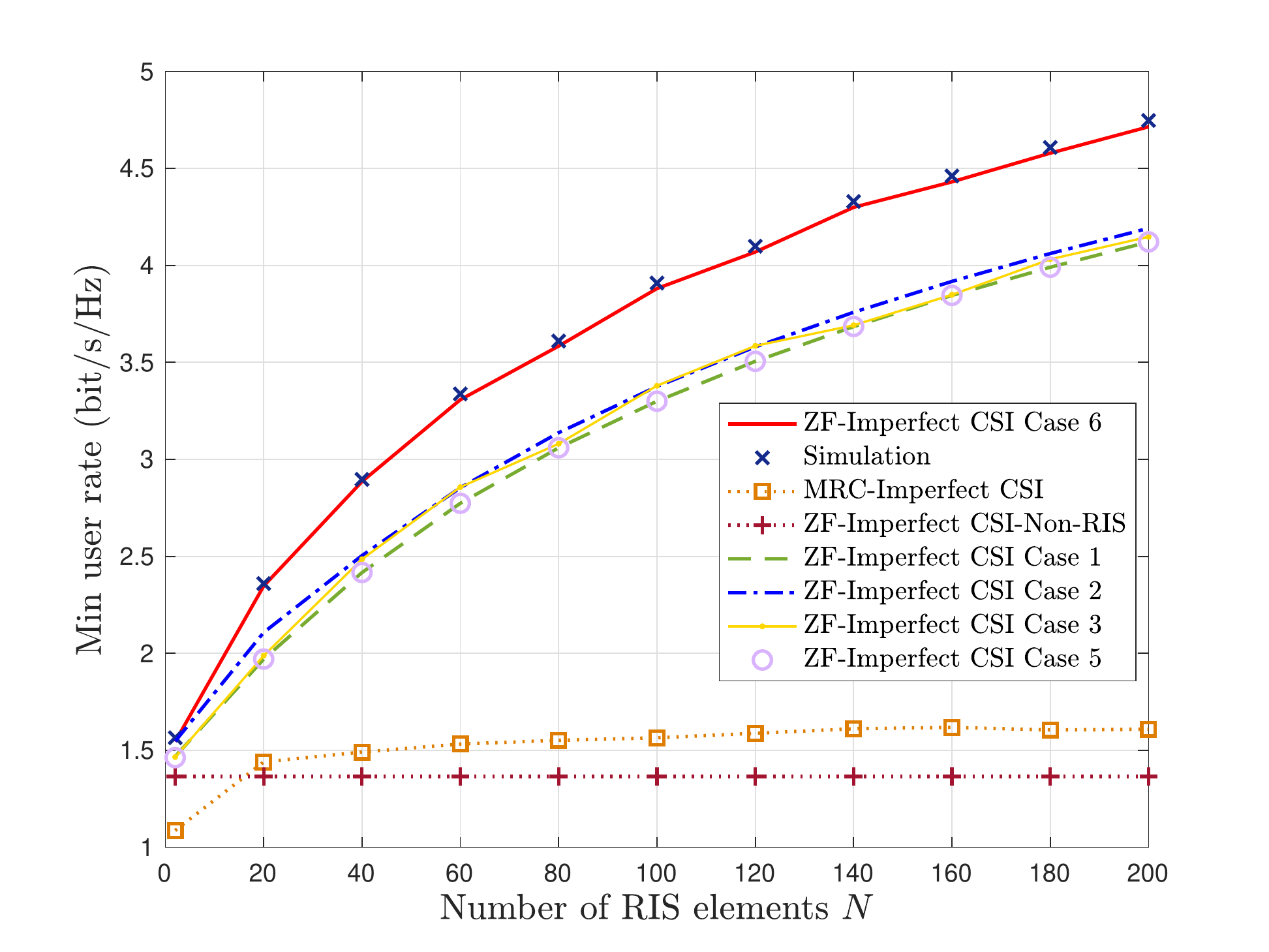}   
		\end{minipage}\label{figure3b}
	}
	
	\caption{Sum user rate and minimum user rate.} 
	\label{figure3} 
\end{figure}

Fig. \ref{figure3a} illustrates the sum user rate. The RIS's phase shifts are designed by solving the Max-Sum problem (\ref{MaxSum}), denoted as \textbf{Case 5}. We also design the RIS's phase shifts based on Case 1 (aligned to user 1), Case 2 (aligned to user 8) and Case 3 (set randomly). Firstly, we can observe some performance loss caused by channel estimation errors. This is because the length of the pilots is $\tau=K=8$, which is very small compared to the large $M$ and $N$. However, the ZF-based perfect and imperfect CSI cases have a similar growth rate (i.e., a nearly constant gap). This is because the channel estimation error $\epsilon_k$ saturates for large $N$ and then does not degrade the scaling order.
Secondly, it is seen that ZF-based systems perform much better than MRC-based and RIS-free systems, especially when $N$ is large. This is consistent with our analytical results. Thirdly, the rate in Case 5 is much higher than that in Case 3. However, a near-optimal performance is achieved by Case 1 and Case 2. Especially, in Case 1 where the RIS phase shifts are aligned to the nearest user, the rate is almost the same as the optimal result.
This is because by aligning the RIS's phase shifts to a user, the rate of this user scales on the order of $\mathcal{O}\left(\log_2\left(MN^2\right)\right)$, while the rates of all other users scale still on the order of $\mathcal{O}\left(\log_2\left(MN\right)\right)$, which corresponds to a large sum user rate for large $M$ and $N$.
Since directly setting $\mathbf{a}_{N}^{H} \mathbf{\Phi} \overline{\mathbf{h}}_{k} = N$ is a very simple and low-complexity approach, aligning the RIS's phase shifts to an arbitrary user is a high-quality sub-optimal solution for practical systems.
Finally, we can again observe the tightness of the lower bound (\ref{rate_much_lowBound}) when $\bf\Phi$ is not optimized.

Fig. \ref{figure3b}  evaluates the minimum user rate. We design the RIS by solving the Max-Min problem (\ref{MaxMin}), denoted as \textbf{Case 6}. We also consider Case 1 (aligned to user 1), Case 2 (aligned to user 8), Case 3 (set randomly), and Case 5 (Max-Sum). It is seen that our optimal design in Case 6 yields better minimum user rates compared with other cases. 
However, despite some performance loss, Cases 1, 2, 3, and 5 also achieve relatively high minimum user rates.
This is because the dominant limitation, namely the multi-user interference, is eliminated. Thus, even the lowest rate grows still on the order of $\mathcal{O}\left(\log_2\left(MN\right)\right)$, which is guaranteed to be high with large $M$ and $N$. 
Meanwhile, we can see that the minimum rates in Case 2 are better than that in Case 1, 3, and 5. This is because in Case 2, the RIS's phase shifts are aligned to the furthest user who has the lowest path-loss factor. Intuitively, compared with Case 1 which aligns the RIS phase shifts to the nearest user, Case 2 is more fair and then achieves a better minimum user rate.

\begin{figure}[htbp]
	\centering    
	
	\subfigure[Trade-off between $M$ and $N$] 
	{
		\begin{minipage}{8cm}
			\centering          
			\includegraphics[scale=0.4]{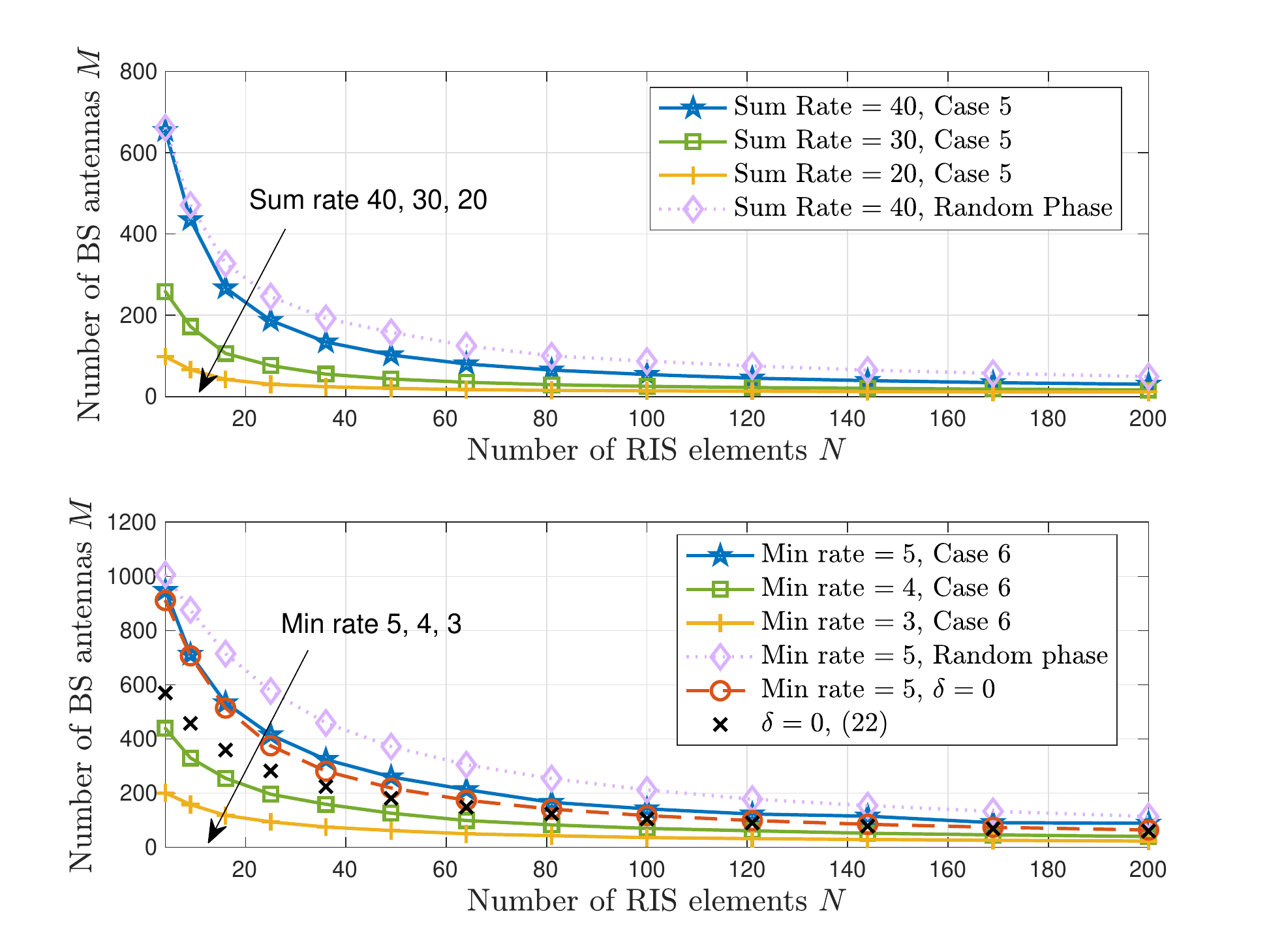}   
		\end{minipage}\label{figure4a}
	}
	\subfigure[Power scaling law, $p=10/N$] 
	{
		\begin{minipage}{7cm}
			\centering      
			\includegraphics[scale=0.4]{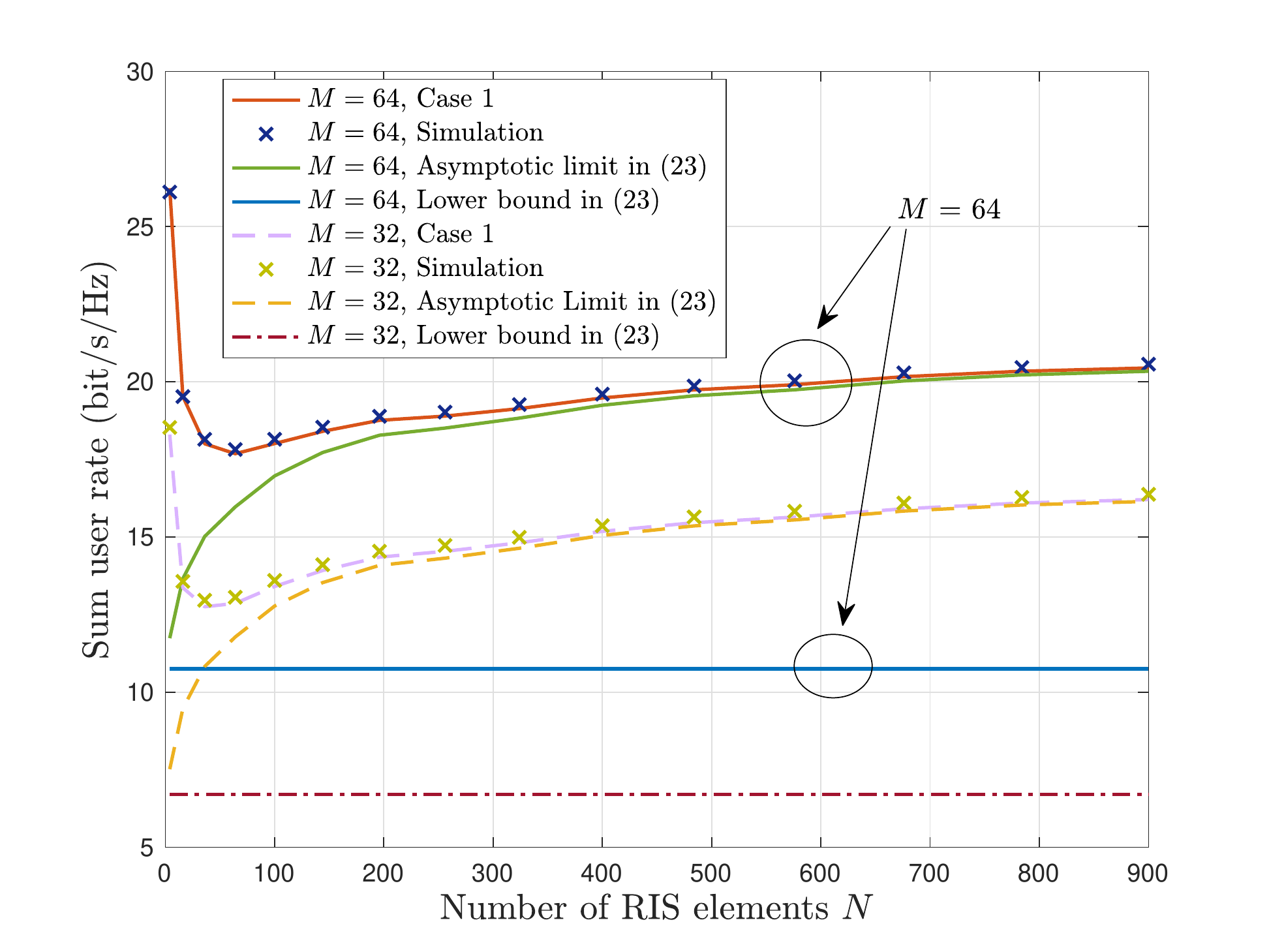}   
		\end{minipage}\label{figure4b}
	}
	
	\caption{$M$ - $N$ trade-off and power scaling law.} 
	\label{figure4}  
\end{figure}

Fig. \ref{figure4a} evaluates the trade-off between $M$ and $N$ with respect to the sum user rate and the minimum user rate, respectively. As expected, in both cases, $ M $ can be reduced inversely proportional to the increase of $ N $ while maintaining a constant rate. Meanwhile, after optimizing $\bf\Phi$, $M$ can be further decreased compared to random phase shifts. We can see that the reduction of $M$ is more obvious if the target is stringent. This comes from the decreasing slope of the logarithm function. Without the RIS, the rate is on the order of $\mathcal{O}\left(\log_2\left(M\right)\right)$, and very large $M$ is needed to achieve a high rate target. However, if the rate is on the order of $\mathcal{O}\left(\log_2\left(MN\right)\right)$, the high data rate target can be met with moderate $M$ but large $N$, since the product $MN$ is very large. 
Besides, when $\delta=0$ (in Corollary \ref{corollary3}), we verify the theoretical relationship (\ref{MN_tradeoff}) by using the path-loss of user 8. 
As can be observed, the derived results are accurate when $N>40$. 


Finally, Fig. \ref{figure4b} validates the derived power scaling law in (\ref{power_LB}),
where the power is scaled proportionally to $p=10/N$. As $N\to\infty$, it is verified that the rate tends to the derived asymptotic limit, and it is larger than the lower bound. Also, by doubling $M$ from $32$ to $64$, we can find a significant increase of the limit. This is because (\ref{power_LB}) is on the order of $\mathcal{O}\left(\log_2\left(M\right)\right)$.


\section{Conclusion}\label{section6}

This work demonstrates that RIS-aided MIMO with ZF detectors is a promising system architecture for many applications. We derive theoretical expressions for the ergodic rate, based on which two low-complexity MM algorithms are proposed to respectively optimize the sum user rate and the minimum user rate. 
We demonstrate that by aligning the RIS phase shifts to a user, the rate scaling order of that user can approach $\mathcal{O}\left(\log_2\left(MN^2\right)\right)$, while the rate scaling order of the other users is guaranteed to be $\mathcal{O}\left(\log_2\left(MN\right)\right)$. Therefore, with a low-complexity RIS design, high system throughput can be realized. We also prove that by increasing $N$, the $M$ required to maintain a constant rate can be reduced inverse proportionally. Besides, we prove that as $N\to\infty$, the transmit power of all users can be scaled proportionally to $p=1/N$ while maintaining high rates.


\begin{appendices}
	
	\section{}\label{App_1}
	Based on the definitions of $\mathbf{H}_1$, $\mathbf{H}_2$, and $\mathbf{D}$, we can expand the channel of user $k$ as follows
	\begin{align}\label{qk_definition}
\mathbf{q}_{k}=\sqrt{\frac{\alpha_{k} \beta \delta}{\delta+1}} \overline{\mathbf{H}}_{2} \mathbf{\Phi} \overline{\mathbf{h}}_{k}+\sqrt{\frac{\alpha_{k} \beta}{\delta+1}} \widetilde{\mathbf{H}}_{2} \mathbf{\Phi} \overline{\mathbf{h}}_{k}+\sqrt{\gamma_{k}} \tilde{\mathbf{d}}_{k}.
	\end{align}

	Firstly, since $\widetilde{\mathbf{H}}_2$ consists of i.i.d. $\mathcal{CN}\left(0,1\right)$ elements, vector $\sqrt{\frac{\alpha_{k} \beta}{\delta+1}} \widetilde{\mathbf{H}}_{2} \mathbf{\Phi} \overline{\mathbf{h}}_{k}$ is comprised of mutually independent elements. 
	Secondly, the elements of vector $\sqrt{\frac{\alpha_{k} \beta}{\delta+1}} \widetilde{\mathbf{H}}_{2} \mathbf{\Phi} \overline{\mathbf{h}}_{k}$ are linear combinations of independent Gaussian random variables. 
	 Therefore, vector $\sqrt{\frac{\alpha_{k} \beta}{\delta+1}} \widetilde{\mathbf{H}}_{2} \mathbf{\Phi} \overline{\mathbf{h}}_{k}$ consists of i.i.d. Gaussian variables, following $ \sqrt{\frac{\alpha_{k} \beta}{\delta+1}} \widetilde{\mathbf{H}}_{2} \mathbf{\Phi} \overline{\mathbf{h}}_{k} \sim \mathcal{CN}\left(\mathbf{0},  N\frac{\alpha_{k} \beta}{\delta+1}  \mathbf{I}_M \right) $. Meanwhile, we have $ \sqrt{\gamma_{k}} \tilde{\mathbf{d}}_{k} \sim \mathcal{CN}\left(\mathbf{0},  \gamma_{k}  \mathbf{I}_M \right) $.
	Since the sum of independent Gaussian vectors is still a Gaussian vector\cite{muirhead2009aspects}, we have
	\begin{align}\label{Gaussian_g_k}
\sqrt{\frac{\alpha_{k} \beta}{\delta+1}} \widetilde{\mathbf{H}}_{2} \mathbf{\Phi} \overline{\mathbf{h}}_{k} +
\sqrt{\gamma_{k}} \tilde{\mathbf{d}}_{k}
\sim \mathcal{CN}\left(\mathbf{0},  \left(N\frac{\alpha_{k} \beta}{\delta+1} + \gamma_{k}\right) \mathbf{I}_M \right).
	\end{align}
	
	Combining (\ref{qk_definition}) and (\ref{Gaussian_g_k}), it is proved that $\mathbf{q}_k$ is a Gaussian distributed vector, where
$ \mathbb{E}\left\{\mathbf{q}_{k}\right\}=\sqrt{\frac{\alpha_{k} \beta \delta}{\delta+1}} \overline{\mathbf{H}}_{2} \mathbf{\Phi} \overline{\mathbf{h}}_{k}, $ and $\operatorname{Cov}\left\{\mathbf{q}_{k}\right\}=\mathbb{E}\{\left(\mathbf{q}_{k}-\mathbb{E}\left\{\mathbf{q}_{k}\right\}\right)\left(\mathbf{q}_{k}-\mathbb{E}\left\{\mathbf{q}_{k}\right\}\right)^{H}\}=\left(N\frac{\alpha_{k} \beta}{\delta+1} + \gamma_{k}\right) \mathbf{I}_M $.

Following a similar procedure, we can derive the distribution of the noise matrix $\frac{1}{\sqrt{\tau p}} \mathbf{N}\mathbf{s}_k$, which is omitted here for brevity.

		\section{}\label{App_2}
	Since the channel $\mathbf{q}_k$ and the noise $\mathbf{N}\mathbf{s}_k$ are Gaussian distributed random variables, the considered observation vector $\mathbf{y}_p^k$ in (\ref{observation_vector}) is consistent with the complex Bayesian linear model \cite[Eq. (15.63)]{kay1993fundamentals}\cite[Lemma B.17]{bjornson2017massive}. Therefore, we can directly apply the results in \cite[Eq. (15.64)]{kay1993fundamentals} and \cite[Eq. (15.67)]{kay1993fundamentals} to obtain the MMSE channel estimate of $\mathbf{q}_k$ and the MSE matrix. In particular, using the distribution in Lemma \ref{lemma1}, we have
	\begin{align}
\hat{\mathbf{q}}_{k}\!=\!\sqrt{\frac{\alpha_{k} \beta \delta}{\delta+1}} \overline{\mathbf{H}}_{2} \mathbf{\Phi} \overline{\mathbf{h}}_{k}
\!+\! \left(\! N \frac{\alpha_{k} \beta}{\delta+1}\! +\! \gamma_{k} \! \right) \! \mathbf{I}_{M} \!
\left( \!\! \left(\! N \frac{\alpha_{k} \beta}{\delta+1}+\gamma_{k}+\frac{\sigma^{2}}{\tau p}\right) \mathbf{I}_{M}\!\right)^{-1}\!\!
\left(\mathbf{y}_{p}^{k}-\sqrt{\frac{\alpha_{k} \beta \delta}{\delta+1}} \overline{\mathbf{H}}_{2} \mathbf{\Phi} \overline{\mathbf{h}}_{k}\right),
	\end{align}
	and
	\begin{align}
\mathbf{M S E}_{k}=\left[\left(\left(N \frac{\alpha_{k} \beta}{\delta+1}+\gamma_{k}\right) \mathbf{I}_{M}\right)^{-1}+\frac{\tau p}{\sigma^{2}} \mathbf{I}_{M}\right]^{-1}.
	\end{align}
	
	After some straightforward simplifications, we can arrive at (\ref{estimate}) and (\ref{MSE_matrix}). Besides, since the channel $\mathbf{q}_k$ is a Gaussian vector, we can obtain that the channel estimate $\hat{\mathbf{q}}_k$ and the estimation error $\mathbf{e}_k$ are independent from each other, due to the orthogonality principle of the MMSE estimator\cite{kay1993fundamentals}.

	\section{}\label{App_3}
	
	To derive the lower bound in (\ref{lb_rate}), we need to calculate the term $\mathbb{E}\{     
	[(\hat{\mathbf{Q}}^{H} \hat{\mathbf{Q}})^{-1}]_{k k}
	\}$, where $\hat{\mathbf{Q}}$ is given in (\ref{hat_Q}). We begin by proving that channel $\hat{\mathbf{Q}}$ is Gaussian distributed.
	\begin{lem}\label{lemma_Gaussian}
\cite{jin2007rank1} A random matrix $\mathbf{X}$ is complex Gaussian distributed as ${\mathbf{X}} \sim \mathcal{C} \mathcal{N}(\mathbf{E}, \mathbf{\Sigma} \otimes \mathbf{\Psi})$, if  $\operatorname{vec}\left(\mathbf{X}^{H}\right) \sim \mathcal{C} \mathcal{N}\left(\operatorname{vec}\left(\mathbf{E}^{H}\right), \boldsymbol{\Sigma} \otimes \boldsymbol{\Psi}\right)$. If ${\mathbf{X}_1} \sim \mathcal{C} \mathcal{N}(\mathbf{E}_1, \mathbf{\Sigma}_1 \otimes \mathbf{\Psi}_1)$ and ${\mathbf{X}_2} \sim \mathcal{C} \mathcal{N}(\mathbf{E}_2, \mathbf{\Sigma}_2 \otimes \mathbf{\Psi}_2)$ are independent distributed, then $\mathbf{X}_1+\mathbf{X}_2 \sim \mathcal{C} \mathcal{N} \left(\mathbf{E}_{1}+\mathbf{E}_{2}, \boldsymbol{\Sigma}_{1} \otimes \boldsymbol{\Psi}_{1}+\boldsymbol{\Sigma}_{2} \otimes \boldsymbol{\Psi}_{2}\right)$.
\end{lem}
	
	For notional brevity, we divide the estimated channel $\hat{\mathbf{Q}}$ into three independent parts $\hat{\mathbf{Q}}=\hat{\mathbf{Q}}_{R I S}+\hat{\mathbf{Q}}_{B S}+\hat{\mathbf{Q}}_{ {noise }}$, where
	\begin{align}\label{QQQ}
\begin{aligned}
& \hat{\mathbf{Q}}_{R I S}^{H} =\sqrt{\frac{\beta \delta}{\delta+1}} \mathbf{H}_{1}^{H} \mathbf{\Phi}^{H} \overline{\mathbf{H}}_{2}^{H}+\sqrt{\frac{\beta}{\delta+1}} \mathbf{\Upsilon} \mathbf{H}_{1}^{H} \mathbf{\Phi}^{H} \widetilde{\mathbf{H}}_{2}^{H}, \\
&\hat{\mathbf{Q}}_{BS}^{H}  =\mathbf{\Upsilon} \mathbf{\Omega}^{1 / 2} \widetilde{\mathbf{D}}^{H}, \\
&\hat{\mathbf{Q}}_{noise}^{H}  =\frac{1}{\sqrt{\tau p}} \mathbf{\Upsilon} \mathbf{S}^H\mathbf{N}^{H}.
\end{aligned}
	\end{align}
	
	Recall that $\widetilde{\mathbf{H}}_{2}$, $\widetilde{\mathbf{D}}$, and $\mathbf{N}$ are composed of i.i.d. Gaussian random variables. By observing (\ref{QQQ}), we can find that each column of matrices $\hat{\mathbf{Q}}_{R I S}^{H}$, $\hat{\mathbf{Q}}_{BS}^{H}$ and $\hat{\mathbf{Q}}_{noise}^{H}$ can be written as a
	linear transformation of mutually independent standard Gaussian random vectors. Therefore, the columns of $\hat{\mathbf{Q}}_{R I S}^{H}$, $\hat{\mathbf{Q}}_{BS}^{H}$, and $\hat{\mathbf{Q}}_{noise}^{H}$ are independent  Gaussian vectors.
	  As a result, after vectorization, the vectors $ \operatorname{vec}    ( \hat{\mathbf{Q}}_{R I S}^{H})        $, $ \operatorname{vec}      (\hat{\mathbf{Q}}_{BS}^{H} )     $, and $\operatorname{vec}    (\hat{\mathbf{Q}}_{noise}^{H} )    $ are still Gaussian distributed. 
	  
	  Next, we derive their mean vector and covariance matrices. First, consider the term $ \operatorname{vec}  (\hat{\mathbf{Q}}_{R I S}^{H}  )      $. Obviously, we have $\mathbb{E}\{\operatorname{vec}(\hat{\mathbf{Q}}_{R I S}^{H})\}=  \operatorname{vec}  (\sqrt{\frac{\beta \delta}{\delta+1}} \mathbf{H}_{1}^{H} \mathbf{\Phi}^{H} \overline{\mathbf{H}}_{2}^{H})$. The covariance matrix is given by
	  \begin{align}\label{Q_RIS_cov}
\begin{array}{l}
\operatorname{Cov}\left\{\operatorname{vec}\left(\hat{\mathbf{Q}}_{R I S}^{H}\right)\right\} \!=\! \mathbb{E} \! \left\{\! \operatorname{v e c}\! \left(\sqrt{\frac{\beta}{\delta+1}} \mathbf{\Upsilon} \mathbf{H}_{1}^{H} \mathbf{\Phi}^{H} \widetilde{\mathbf{H}}_{2}^{H} \mathbf{I}_{M}\right) \! \operatorname{vec} \! \left(\sqrt{\frac{\beta}{\delta+1}} \mathbf{\Upsilon} \mathbf{H}_{1}^{H} \mathbf{\Phi}^{H} \widetilde{\mathbf{H}}_{2}^{H} \mathbf{I}_{M}\right)^{H}\right\} \\
\overset{(c)}{=}\left(\mathbf{I}_{M} \otimes \sqrt{\frac{\beta}{\delta+1}} \mathbf{\Upsilon} \mathbf{H}_{1}^{H} \mathbf{\Phi}^{H}\right) \mathbb{E}\left\{\operatorname{vec}\left(\widetilde{\mathbf{H}}_{2}^{H}\right) \operatorname{vec}\left(\widetilde{\mathbf{H}}_{2}^{H}\right)^{H}\right\}\left(\mathbf{I}_{M} \otimes \sqrt{\frac{\beta}{\delta+1}} \mathbf{\Phi} \mathbf{H}_{1} \mathbf{\Upsilon}\right) \\
=\left(\mathbf{I}_{M} \otimes \sqrt{\frac{\beta}{\delta+1}} \mathbf{\Upsilon} \mathbf{H}_{1}^{H} \mathbf{\Phi}^{H}\right)\left(\mathbf{I}_{M} \otimes \sqrt{\frac{\beta}{\delta+1}} \mathbf{\Phi} \mathbf{H}_{1} \mathbf{\Upsilon}\right) \overset{(d)}{=}\mathbf{I}_{M} \otimes \frac{\beta}{\delta+1} \mathbf{\Upsilon} \mathbf{H}_{1}^{H} \mathbf{H}_{1} \mathbf{\Upsilon},
\end{array}
	  \end{align}
	where $(c)$ utilizes $\operatorname{vec}(\boldsymbol{A} \boldsymbol{B} \boldsymbol{C})=\left(\boldsymbol{C}^{{T}} \otimes \boldsymbol{A}\right) \operatorname{vec}(\boldsymbol{B})$ and $(\boldsymbol{A} \otimes \boldsymbol{B})^{{H}}=\boldsymbol{A}^{{H}} \otimes \boldsymbol{B}^{{H}}$. $(d)$ exploits $(\boldsymbol{A} \otimes \boldsymbol{C})(\boldsymbol{B} \otimes \boldsymbol{D})=(\boldsymbol{A} \boldsymbol{B}) \otimes(\boldsymbol{C D})$ and $\mathbf{\Phi}^H\mathbf{\Phi}=\mathbf{I}_N$. According to (\ref{Q_RIS_cov}) and Lemma \ref{lemma_Gaussian}, the distribution of $\hat{\mathbf{Q}}_{R I S} $ is given by
	\begin{align}
	\hat{\mathbf{Q}}_{R I S} \sim \mathcal{C} \mathcal{N}\left(\sqrt{\frac{\beta \delta}{\delta+1}} \overline{\mathbf{H}}_{2} \mathbf{\Phi} \mathbf{H}_{1}, \mathbf{I}_{M} \otimes \frac{\beta}{\delta+1} \mathbf{\Upsilon} \mathbf{H}_{1}^{H} \mathbf{H}_{1} \mathbf{\Upsilon}\right).
	\end{align}
	
	Similarly, we can obtain the distribution of $\hat{\mathbf{Q}}_{BS} $ and $\hat{\mathbf{Q}}_{noise} $ as
	\begin{align}
&\hat{\mathbf{Q}}_{B S} \sim \mathcal{C} \mathcal{N}\left(\mathbf{0}, \mathbf{I}_{M} \otimes \mathbf{\Omega} \mathbf{\Upsilon}^2\right), \\
&\hat{\mathbf{Q}}_{ {noise }} \sim \mathcal{C} \mathcal{N}\left(\mathbf{0}, \mathbf{I}_{M} \otimes \frac{\sigma^{2}}{\tau p} \mathbf{\Upsilon}^{2}\right).
	\end{align}
	
	Then, using Lemma \ref{lemma_Gaussian} and the property that $\boldsymbol{A} \otimes \boldsymbol{B} + \boldsymbol{A} \otimes \boldsymbol{C}=\boldsymbol{A} \otimes(\boldsymbol{B} + \boldsymbol{C})$, the estimated channel $\hat{\mathbf{Q}}$ is Gaussian distributed as follows
		\begin{align}
		\begin{aligned}
		\hat{\mathbf{Q}} \sim \mathcal{C} \mathcal{N}\left(\sqrt{\frac{\beta \delta}{\delta+1}} \overline{\mathbf{H}}_{2} \mathbf{\Phi} \mathbf{H}_{1}, \mathbf{I}_{M} \otimes\left(\frac{\beta}{\delta+1} \mathbf{\Upsilon} \mathbf{H}_{1}^{H} \mathbf{H}_{1} \mathbf{\Upsilon}+\mathbf{\Omega} \boldsymbol{\Upsilon}^2+\frac{\sigma^{2}}{\tau p} \mathbf{\Upsilon}^{2}\right)\right).
		\end{aligned}
		\end{align}

\begin{lem}\label{lemma_Wishart}
\cite[Definition 5.1]{ratnarajah2003topics} Let $\mathbf{W}=\mathbf{X}^{H} \mathbf{X}$, with $n\times m$ matrix $ {\mathbf{X}} \sim \mathcal{C} \mathcal{N}(\mathbf{E}, \mathbf{I}_n \otimes \mathbf{\Psi}) $. Then, $\mathbf{W}$ follows a complex non-central Wishart distribution with $n$ degrees of freedom, covariance matrix $\mathbf{\Psi}$, and non-centrality parameter $\mathbf{\Sigma}=\mathbf{\Psi}^{-1} \mathbf{E}^{H} \mathbf{E}$, denoted by $\mathbf{W} \sim \mathcal{C} \mathcal{W}_{m}(n, \mathbf{\Psi}, \mathbf{\Sigma})$. Besides, its mean is $\mathbb{E}(\mathbf{W})=n  \mathbf{\Psi}+ \mathbf{\Psi}\mathbf{\Sigma} $\cite[10.3]{muirhead2009aspects}.
In particular, if $ {\mathbf{X}} \sim \mathcal{C} \mathcal{N}(\mathbf{0}, \mathbf{I}_n \otimes \mathbf{\Psi}) $ has zero mean, $\mathbf{W}$ is complex central Wishart distributed, denoted by $\mathbf{W} \sim \mathcal{C} \mathcal{W}_{m}(n, \mathbf{\Psi})$, where $\mathbb{E}(\mathbf{W})=n  \mathbf{\Psi}$ and $\mathbb{E}(\mathbf{W}^{-1})=\frac{1}{n-m}  \mathbf{\Psi}^{-1}$, $n>m$ \cite{tague1994expectations}.
\end{lem}
	
	Since $\hat{\mathbf{Q}}$ is Gaussian distributed, from Lemma \ref{lemma_Wishart}, the product $\hat{\mathbf{Q}}^{H} \hat{\mathbf{Q}}$ follows a complex non-central Wishart distribution as
	\begin{align}\label{non_central_Wishart}
	\hat{\mathbf{Q}}^{H} \hat{\mathbf{Q}}\sim \mathcal{CW}_{K}
\left(M, \mathbf{\Psi}_{RIS},\mathbf{\Sigma}_{RIS}\right),
	\end{align}
	where $\mathbf{\Psi}_{RIS}=\frac{\beta}{\delta+1} \mathbf{\Upsilon} \mathbf{H}_{1}^{H} \mathbf{H}_{1} \mathbf{\Upsilon}+ \mathbf{\Omega}\mathbf{\Upsilon}^2+\frac{\sigma^{2}}{\tau p} \mathbf{\Upsilon}^{2}$ and $\mathbf{\Sigma}_{RIS} = \left(\mathbf{\Psi}_{RIS}\right)^{-1} \frac{\beta \delta}{\delta+1} \mathbf{H}_{1}^{H} \mathbf{\Phi}^{H} \overline{\mathbf{H}}_{2}^{H} \overline{\mathbf{H}}_{2} \mathbf{\Phi} \mathbf{H}_{1}$.
	It has been proved that the non-central Wishart distribution can be closely approximated by a central Wishart distribution\cite{steyn1972approximations}. Therefore, as in \cite{zhang2014power,Cons2012ZF,Cons2015schur}, we approximate the non-central Wishart distribution (\ref{non_central_Wishart}) by a central one with the same first order moment. With Lemma \ref{lemma_Wishart} and (\ref{hk2}), the mean of (\ref{non_central_Wishart}) is given by
	\begin{align}
\begin{aligned}
\mathbb{E}\left\{\hat{\mathbf{Q}}^{H} \hat{\mathbf{Q}}\right\} &=M\left(\frac{\beta}{\delta+1} \mathbf{\Upsilon} \mathbf{H}_{1}^{H} \mathbf{H}_{1} \mathbf{\Upsilon}+ \mathbf{\Omega} \mathbf{\Upsilon}^2+\frac{\sigma^{2}}{\tau p} \mathbf{\Upsilon}^{2}\right)+\frac{\beta \delta}{\delta+1} \mathbf{H}_{1}^{H} \mathbf{\Phi}^{H} \overline{\mathbf{H}}_{2}^{H} \overline{\mathbf{H}}_{2} \mathbf{\Phi} \mathbf{H}_{1} \\
&=M\left(\frac{\beta}{\delta+1} \mathbf{\Upsilon} \mathbf{H}_{1}^{H} \mathbf{H}_{1} \mathbf{\Upsilon}+ \mathbf{\Omega}\mathbf{\Upsilon}^2+\frac{\sigma^{2}}{\tau p} \mathbf{\Upsilon}^{2}\right)+M \frac{\beta \delta}{\delta+1} \mathbf{H}_{1}^{H} \mathbf{\Phi}^{H} \mathbf{a}_{N} \mathbf{a}_{N}^{H} \mathbf{\Phi} \mathbf{H}_{1}.
\end{aligned}
	\end{align}
	Then, the central Wishart distribution with the same mean is given by
	\begin{align}
\hat{\mathbf{Q}}^{H} \hat{\mathbf{Q}} \sim  \mathcal{CW}_{K} \left( M, \frac{\beta}{\delta+1}\mathbf{\Upsilon}\mathbf{H}_{1}^{H} \mathbf{H}_{1} \mathbf{\Upsilon}+ \mathbf{\Omega} \mathbf{\Upsilon}^2+\frac{\sigma^{2}}{\tau p} \mathbf{\Upsilon}^{2}+\frac{\beta \delta}{\delta+1} \mathbf{H}_{1}^{H} \mathbf{\Phi}^{H} \mathbf{a}_{N} \mathbf{a}_{N}^{H} \mathbf{\Phi} \mathbf{H}_{1}\right).
	\end{align}
	Aided by the property of complex central Wishart distribution in Lemma \ref{lemma_Wishart}, we obtain
	\begin{align}\label{wishart_inverse}
\mathbb{E}\left\{\left(\hat{\mathbf{Q}}^{H} \hat{\mathbf{Q}}\right)^{-1}\right\}=\frac{\left(\frac{\beta}{\delta+1} \mathbf{\Upsilon} \mathbf{H}_{1}^{H} \mathbf{H}_{1} \mathbf{\Upsilon}+ \mathbf{\Omega} \mathbf{{\Upsilon}}^2+\frac{\sigma^{2}}{\tau p} \mathbf{\Upsilon}^{2}+\frac{\beta \delta}{\delta+1} \mathbf{H}_{1}^{H} \mathbf{\Phi}^{H} \mathbf{a}_{N} \mathbf{a}_{N}^{H} \mathbf{\Phi} \mathbf{H}_{1}\right)^{-1}}{M-K}.
	\end{align}
	
	The proof is completed by substituting (\ref{wishart_inverse}) into (\ref{lb_rate}).
	
	\section{}\label{App_4}
	Recall that $\mathbf{\Lambda} = \frac{\beta}{\delta+1} \mathbf{\Upsilon} \mathbf{H}_{1}^{H} \mathbf{H}_{1} \mathbf{\Upsilon}+\mathbf{\Omega} \boldsymbol{\Upsilon}^{2}+\frac{\sigma^{2}}{\tau p} \mathbf{\Upsilon}^{2}$.
	It is readily found that $\mathbf{\Lambda}=\mathbf{\Lambda}^H$. 
	Note that we assume the existence of direct links, therefore we have $\mathbf{\Omega}\succ \mathbf{0}$.
	Meanwhile, we have $\frac{\beta}{\delta+1} \mathbf{\Upsilon} \mathbf{H}_{1}^{H} \mathbf{H}_{1} \mathbf{\Upsilon}\succeq \mathbf{0}$, $\mathbf{\Omega} \mathbf{\Upsilon}^{2}\succ \mathbf{0}$, and $\frac{\sigma^{2}}{\tau p} \mathbf{\Upsilon}^{2}\succ\mathbf{0}$. Therefore, we obtain that $\mathbf{\Lambda}\succ\mathbf{0}$, $\mathbf{\Lambda}^{-1}\succ\mathbf{0}$, and $\left(\mathbf{\Lambda}^{-1}\right)^H=\mathbf{\Lambda}^{-1}$. Then, applying the Woodbury's identity and using the fact that $\mathbf{\Lambda}^{-1}$ is positive definite and Hermitian, we have
	\begin{align}\label{inverse_lowerBound}
&\left[\left(\boldsymbol{\Lambda}+\frac{\beta \delta}{\delta+1} \mathbf{H}_{1}^{H} \mathbf{\Phi}^{H} \mathbf{a}_{N} \mathbf{a}_{N}^{H} \mathbf{\Phi} \mathbf{H}_{1}\right)^{-1}\right]_{k k} =\left[\boldsymbol{\Lambda}^{-1}\right]_{k k}-\frac{\frac{\beta \delta}{\delta+1}\left[\boldsymbol{\Lambda}^{-1} \mathbf{H}_{1}^{H} \mathbf{\Phi}^{H} \mathbf{a}_{N} \mathbf{a}_{N}^{H} \mathbf{\Phi} \mathbf{H}_{1} \mathbf{\Lambda}^{-1}\right]_{k k}}{1+\frac{\beta \delta}{\delta+1} \mathbf{a}_{N}^{H} \mathbf{\Phi} \mathbf{H}_{1} \mathbf{\Lambda}^{-1} \mathbf{H}_{1}^{H} \mathbf{\Phi}^{H} \mathbf{a}_{N}} \nonumber\\
&=\left[\boldsymbol{\Lambda}^{-1}\right]_{k k}-\frac{\frac{\beta \delta}{\delta+1}\left|\left[\boldsymbol{\Lambda}^{-1} \mathbf{H}_{1}^{H} \mathbf{\Phi}^{H} \mathbf{a}_{N}\right]_{(k, 1)}\right|^{2}}{1+\frac{\beta \delta}{\delta+1}\left(\mathbf{a}_{N}^{H} \mathbf{\Phi} \mathbf{H}_{1}\right) \mathbf{\Lambda}^{-1}\left(\mathbf{a}_{N}^{H} \mathbf{\Phi} \mathbf{H}_{1}\right)^{H}} \leq\left[\boldsymbol{\Lambda}^{-1}\right]_{k k}.
	\end{align}
	
Substituting (\ref{inverse_lowerBound}) into (\ref{rate}), we can obtain the lower bound in (\ref{rate_much_lowBound}). 

\begin{lem}\label{hkhi}
\cite{zhi2021twotimescale,zhang2014power,ozdogan2019massive} When $N\to\infty$, the product of the LoS components $\overline{\mathbf{h}}_{k}^{H} \overline{\mathbf{h}}_{i}$ is still bounded, unless user $i$ has the same AoA as user $k$. 
\end{lem}

We can respectively calculate the diagonal and non-diagonal elements of $\mathbf{\Lambda}$ as follows
\begin{align}\label{HH_diag}
&\left[   \mathbf{\Lambda}\right]_{(k, k)}=   \left(  N\frac{\alpha_{k}\beta}{\delta+1} + \gamma_{k}+\frac{\sigma^{2}}{\tau p}  \right) \kappa_{k}^2,  \\\label{HH_Ndiag}
&{\left[\mathbf{\Lambda}\right]_{(k, i)}=\frac{\beta}{\delta+1}   \sqrt{\alpha_{k} \alpha_{i}} {\kappa}_{k} \kappa_{i} \overline{\mathbf{h}}_{k}^{H} \overline{\mathbf{h}}_{i}, \quad \forall i \neq k.}
\end{align}
When $N$ is small, due to the small product-distance path loss ${\alpha_{k}}\beta$  and $\sqrt{\alpha_{i}\alpha_{k}}\beta$ compared with $\gamma_{k}$, (\ref{HH_Ndiag}) is much smaller compared with (\ref{HH_diag}). Therefore, $\mathbf{\Lambda}$ can be approximated as a diagonal matrix for small $N$. When $N$ increases, based on Lemma \ref{hkhi}, (\ref{HH_diag}) grows much faster than (\ref{HH_Ndiag}).
Thus,  (\ref{HH_diag}) is still much larger than (\ref{HH_Ndiag}) and we can approximate that $\mathbf{\Lambda}$ is dominated by diagonal elements. Finally, when $N\to\infty$, (\ref{HH_diag}) tends to infinity but (\ref{HH_Ndiag}) does not. Therefore, $\mathbf{\Lambda}$ tends to a diagonal matrix for large $N$. Accordingly, for any $N$, we can approximate $\mathbf{\Lambda}$ as a diagonal matrix $\operatorname{diag}\{ \left[   \mathbf{\Lambda}\right]_{(1, 1)},\dots,\left[   \mathbf{\Lambda}\right]_{(K, K)}   \}$
and then arrive the approximate lower bound in (\ref{loweBound_approx}) by using
$ \left[\boldsymbol{\Lambda}^{-1}\right]_{k k} \approx 
\left( \left[\boldsymbol{\Lambda}\right]_{k k} \right)^{-1}
=\frac{N \frac{\alpha_{k} \beta}{\delta+1}+\gamma_{k}+\frac{\sigma^{2}}{\tau p}} {\left(N \frac{\alpha_{k} \beta}{\delta+1}+\gamma_{k}\right)^{2}} $. Finally, by observing the order of magnitude of the numerator and denominator of the SNR in (\ref{loweBound_approx}), we can find that the numerator is on the order of $\mathcal{O}\left(MN^2\right)$, but the denominator is only on the order of $\mathcal{O}\left(N\right)$. Therefore, the rate is on the order of $\mathcal{O}\left(\log_2\left(MN\right)\right)$. Besides, it can be readily found that $\underline{R_{k}}\left(\mathbf{\Phi}\right)=\underline{\underline{R_{k}}}$ when $\delta=0$. Meanwhile, for an optimal solution $\mathbf{\Phi}^{**}$ and a sub-optimal solution $\mathbf{\Phi}^*$, we have $\underline{R_k}\left( \mathbf{\Phi}^{**} \right) > \underline{R_k}\left( \mathbf{\Phi}^{*} \right) $. Since $\underline{\underline{R_{k}}}$ is independent of $\bf\Phi$, we have $\underline{R_k}\left( \mathbf{\Phi}^{**} \right) - \underline{\underline{R_{k}}}> \underline{R_k}\left( \mathbf{\Phi}^{*} \right) -\underline{\underline{R_{k}}}$, which indicates that the gap between $ \underline{R_k}\left( \mathbf{\Phi} \right)  $ and $ \underline{\underline{R_k}} $ will be enlarged if $ \bf\Phi $ is optimized. In other words, the proposed bound $\underline{\underline{R_{k}}}$ will be tight when we use unoptimized phase shifts.

		\section{}\label{App_8}
		
		\begin{lem}\label{lemma_inverse_passiveDefinite}
			If $\mathbf{X}\succ \mathbf{0}$, $ \left[\mathbf{X}^{-1}\right]_{kk} \geq \frac{1}{\left[\mathbf{X}\right]_{kk}} $. The equality holds only if $\mathbf{X}$  is diagonal.
		\end{lem}
		
		Recall that we have $\mathbf{\Lambda}\succ \mathbf{0}$ and $\mathbf{H}_{1}^{H} \mathbf{\Phi}^{H} \mathbf{a}_{N} \mathbf{a}_{N}^{H} \mathbf{\Phi} \mathbf{H}_{1} \succeq \mathbf{0}$. Using Lemma \ref{lemma_inverse_passiveDefinite} and (\ref{H1}), we have
		\begin{align}\label{upb}
			\begin{array}{l}
		{\left[\left( \mathbf{\Lambda}+\frac{\beta \delta}{\delta+1} \mathbf{H}_{1}^{H} \mathbf{\Phi}^{H} \mathbf{a}_{N} \mathbf{a}_{N}^{H} \mathbf{\Phi} \mathbf{H}_{1}\right)^{-1}\right]_{k k} \geq \frac{1}{\left[ \mathbf{\Lambda}+\frac{\beta \delta}{\delta+1} \mathbf{H}_{1}^{H} \mathbf{\Phi}^{H} \mathbf{a}_{N} \mathbf{a}_{N}^{H} \mathbf{\Phi} \mathbf{H}_{1}\right]_{k k}}} =\frac{1}{[ \mathbf{\Lambda}]_{k k}+\frac{\alpha_{k} \beta \delta}{\delta+1}\left|\mathbf{a}_{N}^{H} \mathbf{\Phi} \overline{\mathbf{h}}_{k}\right|^{2}} \\
			{\geq} \frac{1}{[ \mathbf{\Lambda}]_{k k} + \frac{\alpha_{k} \beta \delta}{\delta+1} N^{2}}
			=\frac{1}{ \left(  N\frac{\alpha_{k}\beta}{\delta+1} + \gamma_{k}+\frac{\sigma^{2}}{\tau p}  \right) \kappa_{k}^2 + \frac{\alpha_{k} \beta \delta}{\delta+1} N^{2}},
			\end{array}
		\end{align}
		where the last inequality using the property that $\mathbf{a}_{N}^{H} \mathbf{\Phi} \overline{\mathbf{h}}_{k} \leq N$ from triangle inequality\cite[(189)]{zhi2021twotimescale}, and the equality holds when $\theta_{n}=-\angle\left\{  \left[\mathbf{a}_N^H\right]_{n} \left[\overline{\mathbf{h}}_k\right]_{n} \right\}, \forall n$.
		
		
		The proof is completed by substituting (\ref{upb}) into (\ref{rate}) with a few additional simplifications.

\section{}\label{App_6}	
To begin with, we give a brief introduction to the optimization under the MM framework\cite{sun2017MM,pan2020multicell}. To maximize a function $g(\mathbf{v})$ based on the MM algorithm, at a point $\mathbf{v}_n$, we need to construct a lower bound $\underline{g}(\mathbf{v}|\mathbf{v}_n)$ satisfying
\begin{align}\label{condition0}
&g(\mathbf{v}_n) = \underline{g}\left(\mathbf{v}_n \mid \mathbf{v}_n \right) ,\\\label{condition1}
&g\left(\mathbf{v} \right)  \geq \underline{g}\left(\mathbf{v} \mid \mathbf{v}_n \right),\\\label{condition2}
&\left.\nabla_{\mathbf{v}} {g}\left(\mathbf{v} \right)\right|_{\mathbf{v}=\mathbf{v}_{n}}=
\left.\nabla_{\mathbf{v}}  \underline{g}\left(\mathbf{v} \mid \mathbf{v}_n \right)  \right|_{\mathbf{v}=\mathbf{v}_{n}}.
\end{align}
Then, we are able to increase the value of the original function from $g(\mathbf{v}_n)$ to $g(\mathbf{v}_{n+1})$ by finding the point $\mathbf{v}_{n+1}$ which maximizes the lower bound $\underline{g}(\mathbf{v}|\mathbf{v}_n)$. Therefore, the success of using the MM algorithm highly relies on the properties of the constructed lower bound.

In the following, we derive a tractable lower bound for $f_k(\mathbf{v})$ which satisfies the above three conditions and can successfully produce a closed-form solution.
We first rewrite $f_{k}(\mathbf{v})$ as
\begin{align}
\begin{array}{l}
f_{k}(\mathbf{v})=\ln \left(1+\frac{\mathbf{v}^{H} \mathbf{B v}}{\mathbf{v}^{H} \mathbf{C}_{k} \mathbf{v}}\right)=-\ln \left(\frac{\mathbf{v}^{H} \mathbf{C}_{k} \mathbf{v}}{\mathbf{v}^{H} \mathbf{C}_{k} \mathbf{v}+\mathbf{v}^{H} \mathbf{B v}}\right)  \\
\quad\;\;\quad=-\ln \left(1-\frac{\mathbf{v}^{H} \mathbf{B v}}{\mathbf{v}^{H} \mathbf{C}_{k} \mathbf{v}+\mathbf{v}^{H} \mathbf{B v}}\right) =-\ln \left(1-\frac{\mathbf{v}^{H} \mathbf{B} \mathbf{v}}{t_{k}}\right)
\triangleq f_{k}(\mathbf{v}, t_k),
\end{array}
\end{align}
where $t_{k}=\mathbf{v}^{H}( \mathbf{C}_{k}+\mathbf{B}) \mathbf{v}>0$. Then, according to \cite[(14)]{sun2017MM} and the composition rule \cite[(3.10)]{boyd2004convex}, $ f_{k}(\mathbf{v}, t_k)$ is jointly convex in $\mathbf{v}$ and $t_k$. Therefore, given a point $(\mathbf{v}_n,t_k^n)$, we can obtain a lower bound of $f_{k}(\mathbf{v}, t_k)$ by using its first-order Taylor expansion, which automatically meets the three conditions needed for MM algorithms. Specifically, we have
\begin{align}\label{fk_Taylor1}
\begin{array}{l}
f_{k}(\mathbf{v},t_k) \geq f_{k}\left(\mathbf{v}_{n}, t_{k}^{n}\right)+\left.\frac{\partial f_{k}(\mathbf{v})}{\partial \mathbf{v}^{T}}\right|_{\mathbf{v}=\mathbf{v}_{n}}\!\!\left(\mathbf{v}-\mathbf{v}_{n}\right)+\left.\frac{\partial f_{k}(\mathbf{v})}{\partial \mathbf{v}^{H}}\right|_{\mathbf{v}^{*}=\mathbf{v}_{n}^{*}}\!\!\left(\mathbf{v}^{*}-\mathbf{v}_{n}^{*}\right)+\left.\frac{\partial f_{k}(\mathbf{v})}{\partial t_{k}}\right|_{t_{k}=t_{k}^{n}}\!\!\left(t_{k}-t_{k}^{n}\right),
\end{array}
\end{align}
where $ \frac{\partial     f_{k}(\mathbf{v}, t_k) }{\partial \mathbf{v}^{T}}=\frac{\mathbf{v}^{H} \mathbf{B}}{t_{k}-\mathbf{v}^{H} \mathbf{B} \mathbf{v}} $,
$ \frac{\partial  f_{k}(\mathbf{v}, t_k)  }{\partial \mathbf{v}^{H}}=\frac{\mathbf{v}^{T} \mathbf{B}^{T}}{t_{k}-\mathbf{v}^{H} \mathbf{B} \mathbf{v}} $, and
$ \frac{\partial    f_{k}(\mathbf{v}, t_k)  }{\partial t_{k}}=-\frac{\mathbf{v}^{H} \mathbf{B v}}{\left(t_{k}-\mathbf{v}^{H} \mathbf{B v}\right)} \frac{1}{t_{k}} $.

Substitute these three partial derivatives into (\ref{fk_Taylor1}) and use $ t_k=\mathbf{v}^{H}( \mathbf{C}_{k}+\mathbf{B}) \mathbf{v} $ and $ t_k^n=\mathbf{v}_n^{H}( \mathbf{C}_{k}+\mathbf{B}) \mathbf{v}_n $.  After some simplifications, we can obtain
\begin{align}\label{fk_bound1}
f_{k}(\mathbf{v}) \geq \operatorname{const1}_{k}+2 \operatorname{Re}\left\{\omega_{k} \mathbf{v}_{n}^{H} \mathbf{B} \mathbf{v}\right\}-\psi_{k} \mathbf{v}^{H}\left(\mathbf{C}_{k}+\mathbf{B}\right) \mathbf{v},
\end{align}
where $\operatorname{const1}_{k}=f_{k}\left(\mathbf{v}_{n}\right)-\frac{\mathbf{v}_{n}^{H} \mathbf{B} \mathbf{v}_{n}}{\mathbf{v}_{n}^{H} \mathbf{C}_{k} \mathbf{v}_{n}}$, and $\omega_{k}$ and $\psi_{k}$ were defined in (\ref{omiga_psi}). Next, according to the inequality in \cite[(26)]{sun2017MM} and the property that $ \mathbf{C}_{k}+\mathbf{B} \preceq \lambda_{\max }\left(\mathbf{C}_{k}+\mathbf{B}\right) \mathbf{I}_N $, we have
\begin{align}\label{vBCv_inequlity}
\mathbf{v}^{H}\left(\mathbf{C}_{k}+\mathbf{B}\right) \mathbf{v} & \leq \mathbf{v}^{H} \lambda_{\max }\left(\mathbf{C}_{k}+\mathbf{B}\right) \mathbf{I}_{N} \mathbf{v}+2 \operatorname{Re}\left\{\mathbf{v}^{H}\left(\left(\mathbf{C}_{k}+\mathbf{B}\right)-\lambda_{\max }\left(\mathbf{C}_{k}+\mathbf{B}\right) \mathbf{I}_{N}\right) \mathbf{v}_{n}\right\} \nonumber\\
&+\mathbf{v}_{n}^{H}\left(\lambda_{\max }\left(\mathbf{C}_{k}+\mathbf{B}\right) \mathbf{I}_{N}-\left(\mathbf{C}_{k}+\mathbf{B}\right)\right) \mathbf{v}_{n}.
\end{align}

Substituting (\ref{vBCv_inequlity}) into (\ref{fk_bound1}) and using the fact that $\mathbf{v}^{H} \lambda_{\max }\left(\mathbf{C}_{k}+\mathbf{B}\right) \mathbf{I}_{N} \mathbf{v}=N \lambda_{\max }\left(\mathbf{C}_{k}+\mathbf{B}\right) $, we can arrive at (\ref{fk_lb}).

\section{}\label{App_7}		
Under the MM algorithm framework, given a point $\mathbf{v}_n$, we want to construct a quadratic form lower bound $\underline{\widetilde{f}}\left(\mathbf{v} \mid \mathbf{v}_n \right)$ of $\widetilde{f}\left(\mathbf{v}\right)$ as follows
\begin{align}\label{para_equation}
\widetilde{f}(\mathbf{v}) \geq \underline{\widetilde{f}}\left(\mathbf{v} \mid \mathbf{v}_n \right) = \widetilde{f}\left(\mathbf{v}_{n}\right)+2 \operatorname{Re}\left\{\mathbf{u}^{H}\left(\mathbf{v}-\mathbf{v}_{n}\right)\right\}+\left(\mathbf{v}-\mathbf{v}_{n}\right)^{H} \mathbf{M}\left(\mathbf{v}-\mathbf{v}_{n}\right),
\end{align}
where $\mathbf{u}$ and $\mathbf{M}$ are two parameters to be decided.

Since condition $ \widetilde{f}(\mathbf{v}_n) = \underline{\widetilde{f}}\left(\mathbf{v}_n \mid \mathbf{v}_n \right)  $ is already satisfied, we next construct parameters $\mathbf{u}$ and $\mathbf{M}$ satisfying conditions (\ref{condition1}) and (\ref{condition2}). We first use condition (\ref{condition2}) to design $\mathbf{u}$. The differential of the left hand side of (\ref{para_equation}) at point $\mathbf{v}_n$ with arbitrary increment $\mathrm{d}\mathbf{v}=\mathbf{v}-\mathbf{v}_n$ is
\begin{align}
\begin{array}{l}
\left.\mathrm{d} \widetilde{f}(\mathbf{v})\right|_{\mathbf{v}=\mathbf{v}_{n}}=-\frac{1}{\mu} \frac{\left.\sum\limits_{k} \mathrm{~d}\left\{\exp \left\{-\mu\left(\operatorname{const}_{k}+2 \operatorname{Re}\left\{\left(\mathbf{f}_{k}^{n}\right)^{H} \mathbf{v}\right\}\right)\right\}\right\}\right|_{\mathbf{v}=\mathbf{v}_{n}}}{\sum\limits_{k} \exp \left\{-\mu\left(\mathrm{const}_{k}+2 \operatorname{Re}\left\{\left(\mathbf{f}_{k}^{n}\right)^{H} \mathbf{v}_{n}\right\}\right)\right\}} =\sum\limits_{k} 2 \operatorname{Re}\left\{l_{k}^{n}\left(\mathbf{f}_{k}^{n}\right)^{H}  \mathrm{d}\mathbf{v} \right\},
\end{array}
\end{align}
where $l_k^n$ was defined in (\ref{lkn}). Next, the differential of the right hand side of (\ref{para_equation}) at point $\mathbf{v}_n$ is
\begin{align}
\left.\mathrm{d} \underline{\widetilde{f}}\left(\mathbf{v} \mid \mathbf{v}_{n}\right)\right|_{\mathbf{v}=\mathbf{v}_{n}}=2 \operatorname{Re}\left\{\mathbf{u}^{H}    \mathrm{d}\mathbf{v}    \right\}.
\end{align}
To satisfy condition (\ref{condition2}), we need $ \sum_{k} 2 \operatorname{Re}\left\{l_{k}^{n}\left(\mathbf{f}_{k}^{n}\right)^{H} \mathrm{d} \mathbf{v}\right\}=2 \operatorname{Re}\left\{\mathbf{u}^{H}  \mathrm{d}\mathbf{v}    \right\}  $, resulting in
\begin{align}\label{u_decided}
\mathbf{u}=\sum_{k} l_{k}^{n} \mathbf{f}_{k}^{n}.
\end{align}

Next, we aim to construct $\mathbf{M}$ using condition (\ref{condition1}). Letting $ \mathbf{v}=\mathbf{v}_{n}+\varrho\left(\tilde{\mathbf{v}}-\mathbf{v}_{n}\right) $, $\varrho \in[0,1]$, and substituting it into (\ref{condition1}), we need
\begin{align}\label{equavient_condition}
\widetilde{f}\left(\mathbf{v}_{n}+\varrho\left(\tilde{\mathbf{v}}-\mathbf{v}_{n}\right)\right) \geq \widetilde{f}\left(\mathbf{v}_{n}\right)+2 \varrho \operatorname{Re}\left\{\mathbf{u}^{H}\left(\tilde{\mathbf{v}}-\mathbf{v}_{n}\right)\right\}+\varrho^{2}\left(\tilde{\mathbf{v}}-\mathbf{v}_{n}\right)^{H} \mathbf{M}\left(\tilde{\mathbf{v}}-\mathbf{v}_{n}\right)
\end{align}
to be satisfied for any $\varrho$ and any $\tilde{\mathbf{v}}$. Since we know that $ {\widetilde{f}}\left(\mathbf{v} \right)  $ and $  \underline{\widetilde{f}}\left(\mathbf{v} \mid \mathbf{v}_n \right)$ have the same value and differential at point $\mathbf{v}_n$, (\ref{condition1}) can now be transformed to the condition that the second-order derivative of the left hand side of (\ref{equavient_condition}) is no smaller than that of the right hand side of (\ref{equavient_condition}) for any $\varrho \in[0,1]$ and any $\tilde{\mathbf{v}}$\cite{zhou2020multicast}. 

Specifically, the second-order derivative of the right hand side of (\ref{equavient_condition}) is given by
\begin{align}\label{secon_deri_right}
\frac{\partial}{\partial \varrho^{2}}\left\{\widetilde{f}\left(\mathbf{v}_{n}\right)+2 \varrho \operatorname{Re}\left\{\mathbf{u}^{H}\left(\tilde{\mathbf{v}}-\mathbf{v}_{n}\right)\right\}+\varrho^{2}\left(\tilde{\mathbf{v}}-\mathbf{v}_{n}\right)^{H} \mathbf{M}\left(\tilde{\mathbf{v}}-\mathbf{v}_{n}\right)\right\}=2\left(\tilde{\mathbf{v}}-\mathbf{v}_{n}\right)^{H} \mathbf{M}\left(\tilde{\mathbf{v}}-\mathbf{v}_{n}\right).
\end{align}
Then, we focus on the left hand side of (\ref{equavient_condition}). Its first-order derivative is
\begin{align}
\begin{aligned}
&\frac{\partial}{\partial\varrho} \widetilde{f}\left(\mathbf{v}_{n}+\varrho\left(\tilde{\mathbf{v}}-\mathbf{v}_{n}\right)\right) = \sum_{k} 2\operatorname{Re}\left\{u_{k}^{n}(\varrho)\left(\mathbf{f}_{k}^{n}\right)^{H}\left(\tilde{\mathbf{v}}-\mathbf{v}_{n}\right)\right\},
\end{aligned}
\end{align}
where
$ u_{k}^{n}(\varrho)=\frac{\exp \{-\mu \tilde{l}_k(\varrho)\}}{\sum\limits_{k} \exp \{-\mu \tilde{l}_k(\varrho)\}} $, 
$ \tilde{l}_k(\varrho)=\operatorname{const}_{k}+2 \operatorname{Re}\left\{\left(\mathbf{f}_{k}^{n}\right)^{H}\left(\mathbf{v}_{n}+\varrho\left(\tilde{\mathbf{v}}-\mathbf{v}_{n}\right)\right)\right\} $, and
$ \frac{\partial \tilde{l}_k(\varrho)}{\partial \varrho}=2 \operatorname{Re}\left\{\left(\mathbf{f}_{k}^{n}\right)^{H}\left(\tilde{\mathbf{v}}-\mathbf{v}_{n}\right)\right\} $. Then, we can compute the second-order derivative as follows
\begin{align}\label{second_1}
\frac{\partial}{\varrho^{2}} \widetilde{f}\left(\mathbf{v}^{n}+\varrho\left(\tilde{\mathbf{v}}-\mathbf{v}_{n}\right)\right)= \sum_{k} 2\operatorname{Re}\left\{\frac{\partial}{\partial\varrho}\left\{u_{k}^{n}(\varrho)\right\}\left(\mathbf{f}_{k}^{n}\right)^{H}\left(\tilde{\mathbf{v}}-\mathbf{v}_{n}\right)\right\},
\end{align}
where
\begin{align}\label{seond_2}
\begin{array}{l}
\frac{\partial u_{k}^{n}(\varrho) }{\partial\varrho}\!=\!-2 \mu \operatorname{Re}\left\{u_{k}^{n}(\varrho)\left(\mathbf{f}_{k}^{n}\right)^{H}\left(\tilde{\mathbf{v}}-\mathbf{v}_{n}\right)\right\}\!+\! \mu u_{k}^{n}(\varrho)\left(\sum\limits_{k} 2 \operatorname{Re}\left\{u_{k}^{n}(\varrho)\left(\mathbf{f}_{k}^{n}\right)^{H}\left(\tilde{\mathbf{v}}-\mathbf{v}_{n}\right)\right\}\!\right).
\end{array}
\end{align}
Substituting (\ref{seond_2}) into (\ref{second_1}), we obtain the second-order derivative as follows
\begin{align}\label{second_order_d}
\begin{array}{l}
\frac{\partial}{\partial \varrho^{2}} \widetilde{f}\left(\mathbf{v}_{n}+\varrho\left(\tilde{\mathbf{v}}-\mathbf{v}_{n}\right)\right)
\\=-\mu \sum_{k} u_{k}^{n}(\varrho)\left(2 \operatorname{Re}\left\{\left(\mathbf{f}_{k}^{n}\right)^{H}\left(\tilde{\mathbf{v}}-\mathbf{v}_{n}\right)\right\}\right)^{2}+\mu\left(\sum_{k} 2 \operatorname{Re}\left\{u_{k}^{n}(\varrho)\left(\mathbf{f}_{k}^{n}\right)^{H}\left(\tilde{\mathbf{v}}-\mathbf{v}_{n}\right)\right\}\right)^{2}.
\end{array}
\end{align}
Define $\mathbf{t}=\tilde{\mathbf{v}}-\mathbf{v}_{n}$. (\ref{second_order_d}) can be rewritten as a quadratic form of $ \bf t $, as follows
\begin{align}
&\frac{\partial}{\partial \varrho^{2}} \widetilde{f}\left(\mathbf{v}_{n}+\varrho\left(\tilde{\mathbf{v}}-\mathbf{v}_{n}\right)\right) =\left[\begin{array}{ll}
\mathbf{t}^{H} & \mathbf{t}^{T}
\end{array}\right]\mathbf{W}\left[\begin{array}{l}
\mathbf{t} \\
\mathbf{t}^{*}
\end{array}\right],
\end{align}
where
\begin{align}\label{W_definition}
&\mathbf{W}\!=\!
&\!\!\!\!-\mu \sum_{k} u_{k}^{n}(\varrho)\left[\begin{array}{l}
\mathbf{f}_{k}^{n} \\
\left(\mathbf{f}_{k}^{n}\right)^{*}
\end{array}\right]\left[\begin{array}{l}
\mathbf{f}_{k}^{n} \\
\left(\mathbf{f}_{k}^{n}\right)^{*}
\end{array}\right]^{H}\!\!\!+\mu\left[\begin{array}{l}
\sum_{k} u_{k}^{n}(\varrho) \mathbf{f}_{k}^{n} \\
\sum_{k} u_{k}^{n}(\varrho)\left(\mathbf{f}_{k}^{n}\right)^{*}
\end{array}\right]\left[\begin{array}{l}
\sum_{k} u_{k}^{n}(\varrho) \mathbf{f}_{k}^{n} \\
\sum_{k} u_{k}^{n}(\varrho)\left(\mathbf{f}_{k}^{n}\right)^{*}
\end{array}\right]^{H}.
\end{align}

Besides, we rewrite the second-order derivative in (\ref{secon_deri_right}) as
\begin{align}
 2\left(\tilde{\mathbf{v}}-\mathbf{v}_{n}\right)^{H} \mathbf{M}\left(\tilde{\mathbf{v}}-\mathbf{v}_{n}\right)=\left[\begin{array}{ll}
\mathbf{t}^{H} & \mathbf{t}^{T}
\end{array}\right]\left[\begin{array}{cc}
\mathbf{M} & 0 \\
0 & \mathbf{M}^{T}
\end{array}\right]\left[\begin{array}{l}
\mathbf{t} \\
\mathbf{t}^{*}
\end{array}\right].
\end{align}

To satisfy condition (\ref{condition1}), according to (\ref{W_definition}), we can choose that $\mathbf{M}\preceq\lambda_{\min }(\mathbf{W}) \mathbf{I}_N$, where
\begin{align}\label{lamda_W}
&\lambda_{\min }(\mathbf{W})\overset{(e)}{\geq}   - \mu \sum_{k} u_{k}^{n}(\varrho) \lambda_{\max }\left(\left[\begin{array}{l}
\mathbf{f}_{k}^{n} \\
\left(\mathbf{f}_{k}^{n}\right)^{*}
\end{array}\right]\left[\begin{array}{l}
\mathbf{f}_{k}^{n} \\
\left(\mathbf{f}_{k}^{n}\right)^{*}
\end{array}\right]^{H}\right)\nonumber\\
&\overset{(f)}{=}   -\mu \sum_{k} u_{k}^{n}(\varrho)\left(\left(\mathbf{f}_{k}^{n}\right)^{H} \mathbf{f}_{k}^{n}+\left(\mathbf{f}_{k}^{n}\right)^{T}\left(\mathbf{f}_{k}^{n}\right)^{*}\right) =-2 \mu \sum_{k} u_{k}^{n}(\varrho)\left\|\mathbf{f}_{k}^{n}\right\|^{2} \overset{(g)}{\geq}-2 \mu \max _{k}\left\|\mathbf{f}_{k}^{n}\right\|^{2},
\end{align}
according to the following properties:
$ (e) $\cite{lutkepohl1997handbook} : For Hermitian matrix $\mathbf{X}$ and rank one Hermitian matrix $\mathbf{T}$, we have $\lambda_{\min }(\mathbf{X}+\mathbf{T}) \geq \lambda_{\min }(\mathbf{X}) + \lambda_{\min }(\mathbf{T})= \lambda_{\min }(\mathbf{X})$. 
$ (f) $ : If $\mathbf{X}$ is rank one, $ \lambda_{\max }\left(\mathbf{X}\right) = \mathrm{Tr}\left\{\mathbf{X}\right\} $.
$ (g) $  : For non-negative vector $ [b_1, b_2, . . . , b_n] $ and $ [c_1, c_2, . . . , c_n] $, if $c_i\in\left(0,1\right)$ and $\sum_{i=1}^{n}c_{i}=1$, then $\sum_{i=1}^{n} c_{i} b_{i} \leq \sum_{i=1}^{n} c_{i}\max _{1 \leq i \leq n} b_{i} = \max _{1 \leq i \leq n} b_{i}$.

Based on (\ref{lamda_W}), we can now construct $ \mathbf{M} =    \left(-2 \mu \max _{k}\left\|\mathbf{f}_{k}^{n}\right\|^{2}\right)     \mathbf{I}_N$. Substituting this $\mathbf{M}$ and
 $\mathbf{u}$ in (\ref{u_decided}) into (\ref{para_equation}) completes the proof.

\end{appendices}

\bibliographystyle{IEEEtran}
\vspace{-6pt}
\bibliography{myref.bib}
\end{document}